\documentclass[preprints,article,accept,pdftex,moreauthors]{Definitions/mdpi} 
\firstpage{1} 
\pubvolume{1}
\issuenum{1}
\articlenumber{0}
\pubyear{2025}
\copyrightyear{2025}
\datereceived{ } 
\daterevised{ } 
\dateaccepted{ } 
\datepublished{ } 
\hreflink{https://doi.org/} 

\usepackage{mathtools}
\usepackage{physics}
\newcommand{\bol}[1]{\boldsymbol{#1}}
\usepackage{wasysym}
\usepackage{soul}

\Title{A time-marching quantum algorithm for simulation of the nonlinear Lorenz dynamics}

\TitleCitation{Title}

\Author{Efstratios Koukoutsis $^{1}$*\orcidA{}, George Vahala $^{2}$*\orcidB{}, Min Soe  $^{3}$\orcidC{}, Kyriakos Hizanidis $^{1}$\orcidD{}, Linda Vahala $^{4}$\orcidE{}, Abhay K. Ram $^{5}$\orcidF{} }

\AuthorNames{Firstname Lastname,  and Firstname Lastname}

\isAPAStyle{%
       \AuthorCitation{Lastname, F., Lastname, F., \& Lastname, F.}
         }{%
        \isChicagoStyle{%
        \AuthorCitation{Lastname, Firstname, Firstname Lastname, and Firstname Lastname.}
        }{
        \AuthorCitation{Koukoutsis, E.; Vahala, G.; Soe, M.; Hizanidis, K.; Vahala, L.; Anhay, A.K.}
        }
}

\address{%
$^{1}$ \quad School of Electrical and Computer Engineering, National Technical University of Athens, Zographou 15780, Greece; stkoukoutsis@mail.ntua.gr (E.K.); kyriakos@central.ntua.gr (K.H.)\\
$^{2}$ \quad Department of Physics, William \& Mary, Williamsburg, Virginia 23187, USA; gvahala@gmail.com\\
$^{3}$ \quad Department of Mathematics and Physical Sciences, Rogers State University
, Claremore, Oklahoma 74017, USA; msoe.rsu@gmail.com\\
$^{4}$ \quad Department of Electrical and Computer Engineering, Old Dominion University, Norfolk, Virginia 23529, USA; lvahala@odu.edu\\
$^{5}$  \quad  Plasma Science and Fusion Center, Massachusetts Institute of Technology, Cambridge, Massachusetts 02139, USA; abhay@psfc.mit.edu}
\corres{Correspondence: stkoukoutsis@mail.ntua.gr (E.K.); gvahala@gmail.com (G.V.)}

\abstract{Simulating  nonlinear classical dynamics on a quantum computer is an inherently challenging task due to the linear operator formulation of quantum mechanics. In this work, we provide a systematic approach to alleviate this difficulty by developing a quantum algorithm that implements the time evolution of a second order time-discretized version of the Lorenz model. The Lorenz model is a celebrated system of nonlinear ordinary differential equations that has been extensively studied in the contexts of climate science, fluid dynamics, and chaos theory. Our algorithm possesses a recursive structure and requires only a linear number of copies of the initial state  with respect to the number of integration time-steps. This provides a significant improvement over previous approaches, while preserving the characteristic quantum speed-up in terms of the dimensionality of the underlying differential equations system, that similar time-marching quantum algorithms have previously demonstrated. Notably, by classically implementing the proposed algorithm, we showcase that it accurately captures the structural characteristics of the Lorenz system, reproducing both regular attractors--limit cycles--and the chaotic attractor within the chosen parameter regime.
}

\keyword{Time-marching quantum algorithm; Recursive structure; Hadamard product; Nonlinear ordinary differential equations; Lorenz system;} 

\begin{document}



\section{Introduction}\label{sec:1}

Quantum computing provides a paradigm shift on how we perceive computation by offering the prospect to accelerate certain computational tasks beyond the capabilities of classical computers. In this direction, there is much interest in seeing how quantum computing and information science can be employed to solve important problems in classical physics, and in particular whether quantum computers can be harnessed to speed-up the simulation of complex, nonlinear dissipative classical systems exhibiting classical turbulence and chaos. Naturally, these systems pose significant challenges for integration into the linear and unitary quantum algorithmic framework, as their dynamics are non-norm preserving and nonlinear.

Although the study of dissipative nonlinear classical systems is still in its infancy, various techniques have been employed to amend  the underlying nonlinear differential equations and corresponding dynamics into the framework of quantum simulation. These include Carleman linerization ~\citep{Liu_2021,Krovi_2023,Sanavio_2024,Wu_2025}, the Koopman-von Neumann (KvN) formulation of classical mechanics \citep{Joseph_2020,Novikau_2025}, second quantization methods ~\citep{Shi_2021,May_2024,Shi_2024}, measurement-based approaches \cite{Andress_2025}, quantum variational algorithms~\citep{Lubasch_2020,Pool_2024,Hafshejani_2024} and quantum time-marching schemes~\citep{Leyton_2008,Lloyd_2020,Gaitan_2020,Tennie_2023,Esmaeilifar_2024}. While each method offers certain advantages, the first three are impeded, both in terms of quantum algorithmic advantage and physics-wise, when applied to chaotic systems
~\citep{Andress_2025,Andrande_1981,Lewis_2024,Lin_2024}.

Carleman and KvN based techniques involve an analytic expansion of the solution to a partial differential equation in terms of an order parameter, thereby embedding the finite-dimensional nonlinear system into an infinite-dimensional linear one \citep{Engel_2021}. By truncating the resulting hierarchy of linear equations, quantum simulation algorithms can then be applied to the finite-dimensional linear approximation. However, intuitively speaking, the effectiveness of such linear embedding techniques depends critically on the presence of the Painlevé property \citep{Weiss_1983}, which ensures the convergence of the analytic expansion. This typically restrict the successful application of linear embeddings to regular dynamics~\citep{Andrande_1981}, such as the one-dimensional Burgers equation~\citep{Liu_2021} and the advection-diffusion equation \citep{Novikau_2025}, both known to possess the Painlevé property. Another issue with these linear embedding techniques is that, while they enable the quantum simulation of non-linear classical systems in specific regimes, the necessary projection into a finite-dimensional space can introduce numerical artifacts. These artifacts are difficult to control or even to eliminate them, hence comprising the accuracy of the simulation~\citep{Lin_2024}. Finally, measurement based and second quantization techniques are dominated by quantum information scrambling for long simulation times ~\citep{May_2024,Andress_2025,Nandy_2025}.

On the other hand, quantum implementations of time-marching schemes have shown promising applications in solving  nonlinear differential equations such as a $1$D  cubic nonlinear ordinary differential equation~\citep{Tennie_2023}, as well as Burgers equation for high speed flows~\citep{Esmaeilifar_2024}.

In this paper, we align with the latter approaches by proposing a quantum implementation of classical explicit time-advancement schemes, aiming to develop a quantum solver capable of accurately capturing the chaotic behavior of the Lorenz system. Despite its simplicity, the Lorenz system remains one of the most extensively studied chaotic systems, exhibiting all the essential features of non-Hamiltonian chaos. It has found widespread applications in fluid dynamics, weather forecasting, and chaos theory, among other areas. To implement the time-advancement of a system of nonlinear differential equations we propose a two-stage approach similarly  with the method presented in~\citep{Leyton_2008}. Specifically, by selecting a temporal discretization scheme, the update of the state vector is expressed as a linear operator acting on an augmented nonlinear state in closed form.
The single-time step evolution employs the  Hadamard transform to prepare the nonlinear state, followed by a block encoded version of the linear time-update operator. A key feature of the proposed quantum solver is the adoption of a quantum re-usage of states~\citep{Esmaeilifar_2024} resulting into a recursive structured  quantum time-marching algorithm that maintains the previously established previously quantum-speeds while requiring only a linear number of state copies of states with respect to the integration steps $N_t$. We demonstrate the effectiveness of our construction for a second-order discretized version of the Lorenz system by reproducing the correct qualitative structure of both chaotic and regular attractors.

The structure of the paper is as follows: Section~\ref{sec:1} introduces the basic equations describing the Lorenz dynamics and the presents the respective first- and second-order discretization schemes. The two-stage single time step evolution is presented at the beginning of Section~\ref{sec:2}, consisting of a nonlinear state preparation followed by a non-unitary linear evolution. To create the nonlinear states we build on the Hadamard product in Sections~\ref{sec:3.1.1} and~\ref{sec:3.1.2} while the adopted block encoding technique for implementing the non-unitary evolution operator is presented in Section~\ref{sec:3.2}. In Section~\ref{sec:3.3} the complete quantum time-marching algorithm is illustrated. Then,  in Section~\ref{sec:3.4} the complexity of the quantum algorithm, both in terms of quantum resource scaling and the number of oracle queries is analyzed. Its recursive and post-selective features are also discussed. Comparison of the classically implemented results with those generated with high order and adaptive $Mathematica$ solver is performed in Section~\ref{sec:4}. Finally, in Section~\ref{sec:5} we summarize our findings and outline directions for future research.

\section{The Lorenz model}\label{sec:2}

The Lorenz model \cite{Lorenz_1963} was first explored in the study of heat conduction in atmospheric physics. It consists of a set of $3$ ordinary differential equations with two quadratic nonlinear terms
and $3$ positive free parameters $ \sigma, \rho, \beta$,
\begin{align}
&\dv{x}{t} = \sigma (y-x), \label{Lorenz system1} \\
&\dv{y}{t} = x(\rho - z) - y, \label{Lorenz system2} \\
&\dv{z}{t} = xy - \beta z. \label{Lorenz system3}
\end{align} 
The constants $\sigma$ and $\rho$ are the system's parameters proportional to the Prandtl number and Rayleigh number. The Lorenz equations~\eqref{Lorenz system1}-~\eqref{Lorenz system3} are dissipative since the volume $V(t)$  in phase space enclosed by some surface $S(t)$  evolves as
\begin{equation}\label{flow field}
\frac{dV(t)}{dt} = \int_{S(t)} {\bol F} \cdot {d^2 {\bol a}} = \int_{V(t)} \nabla \cdot {\bol F} {d^3 {\bol v}} = -(\sigma + 1 + \beta) V(t),
\end{equation}
where the vector field ${\bol F}$ has components $(dx/dt, dy/dt, dz/dt)$ and $d^2\bol a,\,\,d^3\bol v$ are the corresponding differential surface and volume elements.  Thus, there is exponential contraction of the phase space volume to a set of measure zero
\begin{equation}\label{exponential decay of the pahse space}
V(t) = e^{-(\sigma + 1 + \beta)t} V(0) \rightarrow 0  \quad \text{as} \quad t \rightarrow \infty .
\end{equation}
Varying the free parameters in the system leads to an incredibly rich bifurcation phenomena, from periodic limit cycles to chaotic attractors with dimension  less than $3$.
\subsection{Selection of the time-marching scheme}\label{sec:2.1}
Applying  a first order forward Euler finite difference scheme in Equations~\eqref{Lorenz system1} -~\eqref{Lorenz system3} in a temporal domain $[t, t+T]$ we obtain in vector form,
\begin{equation}\label{first order}
\mathbf x_{n+1}=\mathbf x_n+ \delta t \bol f(\mathbf x_n), \quad \mathbf x_n=\begin{bmatrix}
x_n\\
y_n\\
z_n
\end{bmatrix},
\end{equation}
where $\mathbf x_n\equiv\mathbf x(t+n\delta t),\,\, n=1,2,...,N_t,\,\, N_t=T/\delta t$   and  $\bol f$ is a polynomial vector function such that $\bol f(\mathbf x)=\hat{C}_1\mathbf x+\hat{C}_2\mathbf x^{\otimes 2}$ with $\hat{C}_{1,2}$ being the coefficient matrices. By embedding the nonlinear terms of Equation~\eqref{first order} into the state vector $\bol \psi_n^{nl}$ a closed set of linear recursive equations is obtained,
\begin{equation}\label{first order augmented}
\bol \psi_{n+1}=\hat{\mathcal{A}}_1\bol \psi^{nl}_n,
\end{equation}
with 
\begin{equation}\label{state definition 1}
\bol \psi_{n+1}=\begin{bmatrix}
\mathbf x_{n+1}\\
\bol 0_{5\times1}
\end{bmatrix},\quad 
\bol\psi_n^{nl}=\begin{bmatrix}
\mathbf x_n\\
x_ny_n\\
x_nz_n\\
\bol 0_{3\times1}
\end{bmatrix},
\end{equation}
and
\begin{equation}\label{A1matrix}
\hat{\mathcal{A}}_1=\begin{bmatrix}
\hat{A}_1 & 0_{3\times3}\\
0_{5\times3}& 0_{5\times5}
\end{bmatrix},\quad \hat{A}_1=\begin{bmatrix}
1-\sigma \delta t&   \sigma \delta t  & 0 & 0 & 0 \\
 \rho \delta t  & 1 - \delta t  & 0 & 0 & -\delta t\\
0  &  0  & 1 - \beta \delta t  & \delta t  & 0 \\
\end{bmatrix}.
\end{equation}

However, the first order Euler scheme runs into numerical problems for stiff differential equations which are not eliminated even for $\delta t << 1$~\citep{Hafshejani_2024}. These computational inaccuracies lead to inability to reproduce the correct Lorenz dynamics. For instance, the first order Euler method for the parameters  set $\sigma = 10,\,\, \rho =28,\,\, \beta = 0.55$  cannot reproduce the period-$2$ limit cycle as found by standard fourth-order Runge-Kutta solvers and indicated in the bifurcation diagram in Figure~\ref{birfucation diagram}. Since we seek an accurate quantum algorithm for the nonlinear dissipative Lorenz system, we consider here a $2$nd order two-stage Runge-Kutta extension of the previous, simple Euler scheme.

In comparison with Equation~\eqref{first order}, the second order difference scheme employs a predictor-corrector of the form~\citep{Süli_2003},
\begin{align}
\Tilde{\mathbf x}_{n+1}&=\mathbf x_n+ \delta t \bol f(\mathbf x_n),\label{second order1}\\
\mathbf x_{n+1}&=\mathbf x_n+\frac{\delta t}{2}[\bol f(\mathbf x_n)+\bol f(\Tilde{\mathbf x}_{n+1})] \label{second order2}.
\end{align}
Notice that the second order time marching scheme in Equations~\eqref{second order1},~\eqref{second order2} is explicit and can be written in an analogous form with that of Equation~\eqref{first order augmented} as follows,
\begin{equation}\label{second order augmented}
\bol \psi_{n+1}=\hat{\mathcal{A}}_2\bol \psi^{nl}_n,
\end{equation}
where now 
\begin{equation}\label{state definition 2}
\bol \psi_{n+1}=\begin{bmatrix}
\mathbf x_{n+1}\\
\bol 0_{13\times1}
\end{bmatrix},\quad \bol \psi_n^{nl}=\begin{bmatrix}
\mathbf x_n \\
x_n y_n \\
x_n z_n \\
y_n z_n \\
x_n y_n^2 \\
x_n^2 y_n \\
x_n^2 z_n \\
x_n y_n z_n\\
\bol 0_{6\times1}
\end{bmatrix},
\end{equation}
and 
\begin{equation}\label{A2matrix}
\hat{\mathcal{A}}_2=\begin{bmatrix}
\hat{A}_2 & 0_{3\times6}\\
0_{13\times10}& 0_{13\times6}
\end{bmatrix},\quad \hat{A}_2=\begin{bmatrix}
a_1&  a_2 & 0 & 0 & a_5 & 0 & 0 & 0 & 0 & 0  \\
b_1 & b_2 & 0 & 0 & b_5 & b_6 & b_7 & b_8 & 0 & 0 \\
0  &  0  & c_3 & c_4 & 0 & 0& 0& 0 & c_9 & c_{10} 
\end{bmatrix}.
\end{equation}
The corresponding coefficients of matrix $\hat{A}_2$ in Equation~\eqref{A2matrix} read,
\begin{align}
& a_1 = 1 - \sigma \delta t + \frac{\sigma \delta t^2}{2} (\rho + \sigma), 
\quad a_2 = \sigma \delta t \left(1- \frac{\delta t}{2} - \frac{\sigma \delta t}{2} \right ),  \quad a_5 = - \frac{\sigma \delta t^2}{2}, \label{a1}\\
& b_1 = \rho \delta t \left( 1- \frac{\delta t}{2} - \frac{\sigma \delta t}{2}   \right ), \quad b_2 = 1-\delta t + \frac{\delta t^2}{2} 
+ \frac{\rho \sigma \delta t^2}{2},  \quad b_5 = \frac{\delta t}{2} \left [-1 + \delta t  - (1- \beta \delta t) (1-\sigma \delta t) \right ],\\
&b_6 = - \frac{\sigma \delta t^2 (1-\beta \delta t )}{2}, \quad b_7 = - \frac{\sigma \delta t^3}{2}, \quad 
b_8 = -\frac{\delta t^2 (1-\sigma \delta t)}{2}, \\
& c_3 = 1 - \beta \delta t + \frac{\beta^2 \delta t^2}{2}, \quad c_4 = \frac{\delta t}{2} \left [ 1 - \beta \delta t + (1-\delta t)
(1-\sigma \delta t)  +  \sigma \rho \delta t^2 \right ], \\
& c_9 = - \frac{\delta t^2}{2} (1-\sigma \delta t), \quad  c_{10} = - \frac{\sigma \delta t^2}{2}.\label{c9-c10}
\end{align}
As expected, these coefficient agree with the Euler $1$st order coefficients in Equation~\eqref{A1matrix}, up to terms of $O(\delta t^2)$.

In the next section, the building blocks of the quantum  algorithm  for implementing the second order scheme in Equation~\eqref{second order augmented} are presented.

\section{The Quantum Algorithm}\label{sec:3}
Based on Equations~\eqref{first order augmented} 
or~\eqref{second order augmented}, the single time step evolution $\bol \psi_n \to \bol \psi_{n+1}$ involves the operations outlined in Table~\ref{single step algorithm}.
\begin{table}[H] 
\caption{Operations required for a single time step update of Equations~\eqref{first order augmented} or~\eqref{second order augmented}. \label{single step algorithm}}
\begin{tabularx}{\textwidth}{L}
\toprule
\textbf{Single step evolution  $n\to n+1$.}	\\
\midrule
$1$. Initial state $\leftarrow$ $\bol \psi_n$	\\
$2$. Preparation of the non-linear state $\bol \psi_n^{nl}$,  $\bol f(\bol \psi_n)=\bol \psi^{nl}_n$\\
$3$. Evolution with the time-advancement matrix $\hat{\mathcal{A}}$, $\bol\psi_{n+1}=\hat{\mathcal{A}}\bol\psi_n^{nl}$ \\
\bottomrule
\end{tabularx}
\end{table}
\noindent 
From a classical computing perspective, the single time-step evolution shown in Table~\ref{single step algorithm} is explicit and straightforward, however
 its implementation on a quantum computer requires careful treatment. This is because steps $2$ and $3$ are not compatible with the linear and unitary framework of quantum computing. From this point onward we adopt the Dirac bra-ket notation $\ket{\bol\psi}$ to indicate the normalized quantum variant of the classical state $\bol \psi$. For instance, the first-order discretized classical state $\bol\psi_n$ from Equation~\eqref{state definition 1} is encoded as a $3$-qubit normalized quantum state  $\ket{\bol\psi_n}$ as follows, 
\begin{equation}\label{Dirac notation}
\ket{\bol\psi_n}=\frac{x_n\ket{0}+y_n\ket{1}+z_n\ket{2}}{\sqrt{x^2_n+y_n^2+z^2_n}},
\end{equation}
with the binary representation of the basis states $\ket{i}$ given by $\ket{0}=\ket{000}_b,\,\,\ket{1}=\ket{001}_b,\,\, \ket{2}=\ket{010}_b$, etc. Thus, since  the linear matrix $\hat{\mathcal{A}}$ is non-unitary, $\mel{\bol \psi}{\hat{\mathcal{A}}^\dagger\hat{\mathcal{A}}}{\bol \psi}\neq\braket{\bol \psi}$,  and the mapping $\ket*{\bol \psi_n}\to\ket{\bol\psi^{nl}_n}$ is nonlinear it is necessary to express these operations within the admissible linear and unitary quantum framework.

Additionally to the previously discussed challenges, extending the single time step procedure to consecutive applications on a quantum computer is not straightforward. This stems from the fact that to prepare the nonlinear state $\ket*{\bol \psi^{nl}}$ one must create 
multiple copies of the initial state $\ket{\bol \psi_n}$,  i.e., this initial state can be quantum prepared and is completely known.

We address 
these issues explicitly by presenting the quantum blocks for single step evolution and then binding them into a complete quantum algorithmic multi-step evolution.

\subsection{Preparation of the nonlinear states}\label{sec:3.1}
The corresponding nonlinear states $\bol\psi^{nl}_n$ for the first- and second-order discretized Lorenz system in Equations~\eqref{state definition 1} and~\eqref{state definition 2}, exhibit polynomial nonlinearities with respect to the amplitudes $x_n, y_n, z_n$ of the $\bol \psi_n$ state. Therefore, the first step is to determine how to implement on a quantum computer the following nonlinear transformation of the amplitudes $\psi_i$, for a $n$-qubit quantum state $\ket{\psi}=\sum_{i=0}^{2^n-1}\psi_i\ket{i}$, 
\begin{equation}\label{general nonlinear transformation}
\ket{\psi}\to\sum_{i} P_i(\psi_i)\ket{i},\quad P_i(\psi_i)=\sum_{j=1}^{K}a_j\psi_i^j,
\end{equation}
where $a_j$ are the coefficients and $K$ the degree of polynomials $P_i(\psi_i)$.

In the following section we describe the implementation procedure of the nonlinear transformation in Equation~\eqref{general nonlinear transformation} using the Hadamard product. Then, we showcase how this approach can serve as a foundation for constructing arbitrary polynomially nonlinear states, and in particular, those arising in the Lorenz system.

\subsubsection{The Hadamard product}\label{sec:3.1.1}
\begin{figure}[H]
\centering
\includegraphics[width=0.4\linewidth]{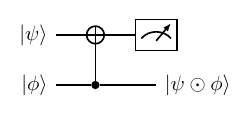}
    \caption{Quantum circuit implementation of Hadamard product $\ket{\psi}\odot\ket{\phi}$ between single-qubit states. A projection measurement operator $\hat{P}=\ket{0}\bra{0}\otimes\hat{1}_{2\times2}$ is applied to the first register for a successful  implementation.}
    \label{Hadamard product circuit}
\end{figure}

The Hadamard product between two $n$ and $m$ qubit states $\ket{\bol\psi}$ and  $\ket{\bol\phi}$ respectively, is defined as follows,
\begin{equation}\label{Hadamard product}
\ket{\bol\psi}\odot\ket{\bol\phi}\equiv\ket{\bol\psi\odot\bol\phi}=\sum_
{i=0}^{2^k-1}\psi_i\phi_i\ket{i},\quad k=max\{n,m\}.
\end{equation}
Boldface quantum states denote multi-qubit states, while non-boldface braket refer to single-qubit states.
Quantum implementation of the Hadamard product~\eqref{Hadamard product} for the case of single-qubit states was 
first presented in~\citep{Pasquinucci_1998}  and is illustrated in Figure~\ref{Hadamard product circuit}.
As depicted in Figure~\ref{Hadamard product circuit}, successful implementation of the Hadamard product is conditioned on the measurement of the $0$-bit in the first register. From here and after, the normalization factor in the  successful post-measurement outcome will be omitted.  For $\ket{\psi}=\ket{\phi}$, we then obtain $\ket{\psi\odot\psi}=\psi_0^2\ket{0}+\psi_1^2\ket{1}$ after tracing out the ancillary $\ket{0}$ state.

\begin{figure}[H]
\centering
\subfloat[\centering]{\includegraphics[width=\linewidth]{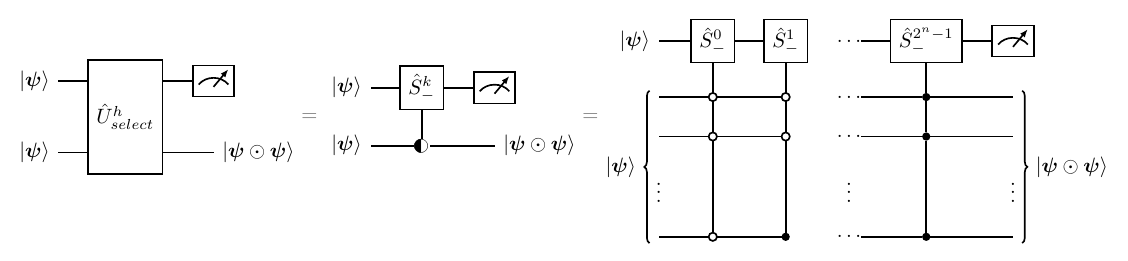}
}
\hfill
\subfloat[\centering]{\includegraphics[]{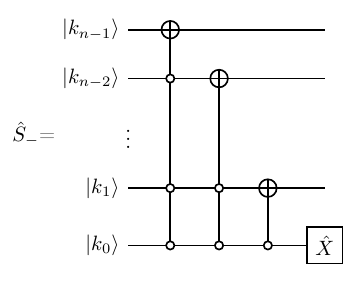}
}
\caption{Quantum circuits for the Hadamard product between two $n$-qubit states. (\textbf{a}) Schematic implementation of the $\hat{U}_{select}^h$ operator. The symbol $\LEFTcircle$ denotes a uniform quantum multiplex gate \citep{Mottonen_2004,Bergholm_2005}. (\textbf{b}) Quantum circuit implementation of the $\hat{S}_{-}$ operation acting on the the $\ket{k}$ basis expressed in its binary form, $\ket{k}=\ket{k_{n-1}k_{n-2}\hdots k_1k_0}_b$. The $\hat{X}$ gate is the Pauli-x matrix.\label{multi Hadamard product circuit}}
\end{figure}

Generalizing the implementation in Figure~\ref{Hadamard product circuit}, the Hadamard product between two $n$-qubit states requires - instead of a single Controlled-NOT (CNOT) gate - the introduction of a multi-control variant $\hat{U}_{select}^h$ through powers of the shift operator $\hat{S}_{-}$,
\begin{equation}\label{Uselecet Hadamard}
\hat{U}_{select}^h=\sum_{k=0}^{2^n-1}\hat{S}^k_{-}\otimes\ket{k}\bra{k},\quad \hat{S}^0_{-}=\hat{1},\quad \hat{S}_{-}\ket{k}=\ket{k-1}.
\end{equation}
Indeed, with operator $\hat{U}_{select}^h$ acting on the $2n$-qubits composite state $\ket{\bol\psi}\otimes\ket{\bol\phi}$ one obtains,
\begin{align}
\hat{U}_{select}^h(\ket{\bol\psi}\otimes\ket{\bol\phi})&=\sum_{ijk}\psi_i\phi_j \hat{S}_{-}^k\ket{i}\otimes\braket{k}{j}\ket{k}=\sum_{ij}\psi_i\phi_j\ket{i-j}\ket{j}\\
&=\sum_{i=j}\psi_i\phi_i\ket{0}\ket{i}+\sum_{i\neq j}\psi_i\phi_j\underbrace{\ket{i-j}}_{\neq\ket{0}}\ket{j}.\label{extra}
\end{align}
From Equation~\eqref{extra}, a successful measurement in the first register with respect to the $0$-bit, provides the Hadamard product transformation between two $n$-qubit states. Figure~\ref{multi Hadamard product circuit} presents the schematic quantum circuit implementations of the  $\ket{\bol\psi\odot\bol\psi}$ transformation in terms of powers of the shift operator $\hat{S}^k_{-}$ along with the corresponding circuit implementation for the $\hat{S}_{-}$.
\noindent Decomposition of the $\hat{S}_{-}$ operator (Figure~\ref{multi Hadamard product circuit}) into CNOTs and single-qubit gates scales as $\textit{O}(n^2)$~\citep{Barenco_1995}. Therefore, the implementation cost of the full sequence of $\hat{S}^k$ gates in Figure~\ref{multi Hadamard product circuit} scales as $\textit{O}(n^2 2^{n+1})$. Given that the $n$-qubit uniformly controlled operations can be implemented within $\textit{O}(2^n)$  elementary gates~\citep{Plesch_2011}, the overall implementation cost for the $\hat{U}_{select}^h$ which realizes the $n$-qubit Hadamard product 
is $\textit{O}[2^n(2n^2+1)]$.

Extension of the Hadamard product to an $N$-tuple of the $n$-qubit state $\ket{\bol\psi}$,
\begin{equation}\label{multi Hadamard product}
\ket*{\underbrace{\bol\psi\odot\bol\psi\odot\ldots\odot\bol\psi}_{N \,\,\text{times}}}\equiv\ket*{\odot^N\bol\psi}=\sum_{i=0}^{2^n-1}\psi^N_i\ket{i},\quad N\geq2,
\end{equation}
can be realized through $N$ copies of the $\ket{\bol\psi}$ acting with the $\hat{U}_h^N$ operator as in 
Equation~\eqref{Uselecet Hadamard},
\begin{equation}\label{U Hadamard gate}
\hat{U}_h^N=\sum_{k=0}^{2^n-1}\underbrace{\hat{S}^k_{-}\otimes\hdots\otimes\hat{S}^k_{-}}_{N-1\,\,\text{times}}\otimes\ket{k}\bra{k}.
\end{equation}
Notice that for $N=1$, $\hat{U}_h=\hat{1}_{2^n\times 2^n}$ and for $N=2$, $\hat{U}_h^2=\hat{U}_{select}^h$ from Equation~\eqref{Hadamard product}. Based on the previous discussion, operator $\hat{U}_h^N$ can be implemented within $\textit{O}[2^n(2n^2N+1)]$ single-qubit and CNOT gates.

Finally, we adopt the following convection in terms of notation
\begin{equation}\label{notation convection for multi Hadamard}
\hat{U}^k_h\ket*{\bol\psi^{\otimes l}}\equiv\ket*{\odot^k\bol\psi},\quad k\leq l,
\end{equation}
indicating that the $\hat{U}^k_h$ operator acts locally on the first $k$ copies from the $l$-fold of the $\ket{\bol\psi}$ with rest of the $l-k$ states omitted to keep the notation uncluttered. 

\subsubsection{Implementing the Lorenz nonlinear states}\label{sec:3.1.2}
Having established all the essential elements about the Hadamard product, the nonlinear transformation in Equation~\eqref{general nonlinear transformation} can be implemented by employing the Hadamard product in conjunction with the Linear Combination of Unitaries (LCU) method \citep{Long_2006}. Specifically, using the definition in Equation~\eqref{U Hadamard gate}, the nonlinear polynomial transformation in Equation~\eqref{general nonlinear transformation} can be expressed as a weighted sum of Hadamard  $\hat{U}_h^j$ gates~\citep{Holmnes_2023},
\begin{equation}\label{nonlinear operator}
\hat{U}_{nl}=\sum_{j=1}^K a_j\hat{U}_h^j,\quad a_j>0.
\end{equation}
Acting with $\hat{U}_{nl}$ on the state $\ket{\bol\psi^{\otimes K}}$ and using the definitions in Equations~\eqref{multi Hadamard product} and ~\eqref{notation convection for multi Hadamard} we obtain,
\begin{equation}\label{U nonlinear}
\hat{U}_{nl}\ket{\bol\psi^{\otimes K}}=\sum_{j=1}^K a_j\ket{\odot^j\bol\psi}=\sum_{j=1}^K\sum_{i=0}^{2^n-1}a_j\psi^j_i\ket{i},
\end{equation}
which coincides with the transformation of Equation~\eqref{general nonlinear transformation}. 

\begin{figure}[H]
\centering
\subfloat[\centering]{\includegraphics[width=0.5\linewidth]{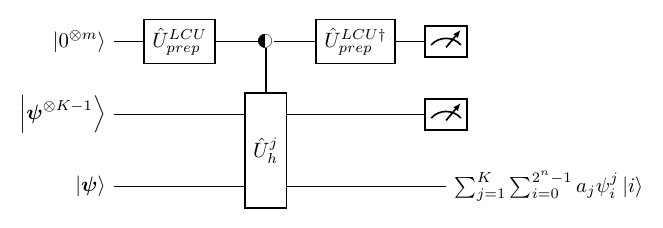}
  }
  \subfloat[\centering]{\includegraphics[width=0.5\linewidth]{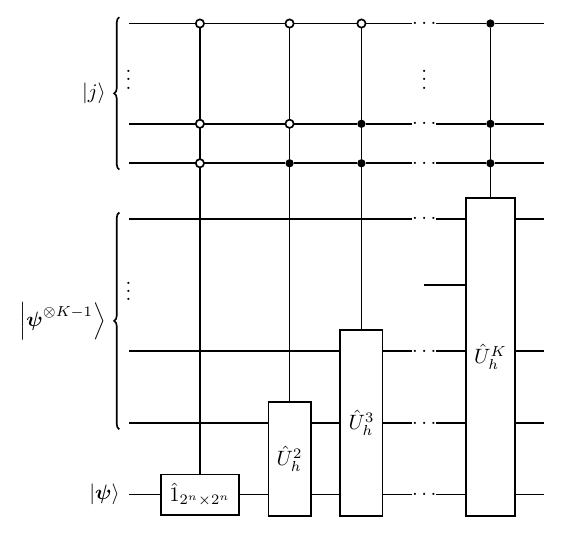}
  }
    \caption{Quantum circuit implementation of the $\hat{U}_{nl}$ operator of Equation~\eqref{U nonlinear} mediating the nonlinear polynomial transformation in Equation~\eqref{general nonlinear transformation}. (\textbf{a}) Schematic implementation of $\hat{U}_{nl}$. (\textbf{b}) Explicit implementation of the uniform multiplexed operation  $\hat{U}^{LCU}_{select}$ in Equation~\eqref{uselect}. The decomposition for each of the $nj$-qubit $\hat{U}^j_h$ operations can be derived from the respective one in Figure~\ref{multi Hadamard product circuit}.}
    \label{non linear operator circuit}
\end{figure}

Following the LCU implementation lemma \citep{Childs_2012}, we define the following unitary operators:
\begin{align}
&\hat{U}^{LCU}_{prep}\ket{0^{\otimes m}}=\frac{1}{\sqrt{a}}\sum_{j=1}^{K}\sqrt{a_j}\ket{j},\quad m=\log_2{K},\quad a=\sum_{j=1}^Ka_j,\\
&\hat{U}^{LCU}_{select}=\sum_{j=1}^K \ket{j}\bra{j}\otimes\hat{U}^j_h,\label{uselect}
\end{align}
which facilitate the circuit implementation of operator $\hat{U}_{nl}$ from Equation~\eqref{nonlinear operator}, as  illustrated in Figure~\ref{non linear operator circuit}.

\begin{figure}[H]
\centering
\includegraphics[]{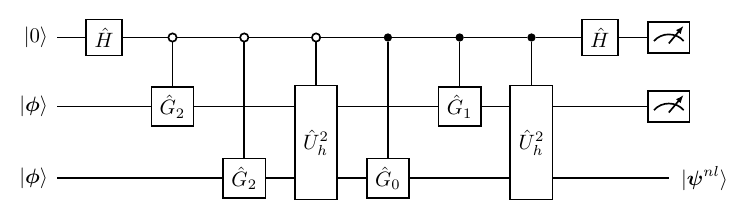}
    \caption{Explicit quantum circuit preparation of the nonlinear state $\ket*{\bol\psi^{nl}}$ in Equation~\eqref{tensor product non linear}. Implementation of $\hat{U}_h^2$ operator has been presented at Figure~\ref{multi Hadamard product circuit}.}
    \label{explicit nonlinear state preparation circuit}
\end{figure}

Introducing into the sum in Equation~\eqref{nonlinear operator} unitary multi-CNOT gates $\hat{G}$ that alternate the amplitudes $\psi_i$ within the state $\ket{\bol\psi}$, we can prepare any type of polynomial nonlinear state. For example, suppose we aim to implement the following nonlinear $2$-qubit state,
\begin{equation}\label{tensor product non linear}
\ket*{\bol\psi^{nl}}=\ket{\psi}\otimes\ket{\psi}=\begin{bmatrix}
\psi_0^2\\
\psi_0\psi_1\\
\psi_1\psi_0\\
\psi_1^2
\end{bmatrix}.
\end{equation}
This type of term arises in the first order difference scheme, Equation~\eqref{state definition 1}.
Then, the nonlinear state~\eqref{tensor product non linear} translates into
\begin{equation}
\ket*{\bol\psi^{nl}}=\hat{G}_2\ket{\bol\phi}\odot\hat{G}_2\ket{\bol\phi}+\hat{G}_1\ket{\bol\phi}\odot\hat{G}_0\ket{\bol\phi},\quad \ket{\bol\phi}=\ket{0}\ket{\psi},
\end{equation}
where the operators $\hat{G}$ are,
\begin{equation}
\hat{G}_0=\begin{bmatrix}
0 &1 &0&0\\
1&0&0&0\\
0&0&1&0\\
0&0&0&1
\end{bmatrix},\quad \hat{G}_1=\begin{bmatrix}
0&0&1&0\\
0&0&0&1\\
1&0&0&0\\
0&1&0&0
\end{bmatrix},\quad \hat{G}_2=\begin{bmatrix}
1&0&0&0\\
0&0&0&1\\
0&0&1&0\\
0&1&0&0
\end{bmatrix}.
\end{equation}
Therefore, the implementation operator $\hat{U}_{nl}$ is given by,
\begin{equation}\label{Final 1 example}
\hat{U}_{nl}=\hat{U}_h^2\hat{G}^{\otimes 2}_2+\hat{U}_h^2(\hat{G}_1\otimes\hat{G}_0).
\end{equation}
Quantum circuit implementation of $\hat{U}_{nl}$ in Equation~\eqref{Final 1 example} follows the same employment of the LCU method as in Figure~\ref{non linear operator circuit} and is explicitly depicted in Figure~\ref{explicit nonlinear state preparation circuit}.
\noindent The $\hat{G}$ gates alternating the amplitudes can be explicitly implemented though the Gray code \cite{Nielsen_2010}. In particular, $\hat{G}_0=\ket{0}\bra{0}\otimes\hat{X}+\ket{1}\bra{1}\otimes\hat{1}_{2\times2}$, $\hat{G}_1=\hat{X}\otimes\hat{X}$ and $\hat{G}_2=\hat{1_{2\times2}}\otimes\ket{0}\bra{0}+\hat{X}\otimes\ket{1}\bra{1}$.

Finally, the full nonlinear state $\ket*{\bol\psi^{nl}_n}$ for the first order discretized Lorenz model in Equation~\eqref{state definition 1} can be realized though the operator,
\begin{equation}\label{nonliner operator Lorenz 1}
\hat{U}_{nl}=\hat{U}_h^1+ \hat{U}_h^2(\hat{G}_1\otimes\hat{G}_0),\quad \hat{U}_h^1=\hat{1}_{8\times8},
\end{equation}
with
\begin{equation}
\hat{G}_0=\begin{bmatrix}
0&0&0&1&0&0&0&0\\
0&0&0&0&1&0&0&0\\
0&0&0&0&0&1&0&0\\
1&0&0&0&0&0&0&0\\
0&1&0&0&0&0&0&0\\
0&0&1&0&0&0&0&0\\
0&0&0&0&0&0&1&0\\
0&0&0&0&0&0&0&1
\end{bmatrix},\quad \hat{G}_1=\begin{bmatrix}
0&0&0&1&0&0&0&0\\
0&0&0&0&1&0&0&0\\
0&0&0&0&0&1&0&0\\
0&1&0&0&0&0&0&0\\
1&0&0&0&0&0&0&0\\
0&0&0&0&0&0&1&0\\
0&0&1&0&0&0&0&0\\
0&0&0&0&0&0&0&1
\end{bmatrix}.
\end{equation}
Therefore, implementing the $\hat{U}_{nl}$ from Equation~\eqref{nonliner operator Lorenz 1} to generate the nonlinear state $\ket*{\bol\psi^{nl}_n}$ in Equation~\eqref{state definition 1}, as well as its counterpart for the second-order discretization in Equation~\eqref{state definition 2}, involves a straightforward extension of replacing the two $2$-qubit copies in Figure~\ref{explicit nonlinear state preparation circuit} with two $3$-qubit and three $4$-qubit copies of state $\ket{\bol\psi_n}$, respectively.

\subsection{Block encoding of non-unitary matrix $\hat{\mathcal{A}}$}\label{sec:3.2}
\begin{figure}[H]
\centering
\includegraphics[width=0.4\linewidth]{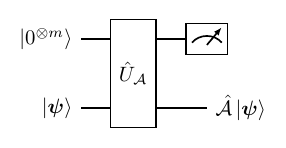}
    \caption{Quantum circuit implementation of the non-unitary operator $\hat{\mathcal{A}}$ through the general block encoding $\hat{U}_{\mathcal{A}}$ in Equations~\eqref{block encoded},~\eqref{action of block encoded}. }
    \label{block encoding circuit}
\end{figure}

Implementing a non-unitary  $2^n\times 2^n$ matrix $\hat{\mathcal{A}}$ requires a $(m,a)$-block encoding of the form,
\begin{equation}\label{block encoded}
\hat{U}_{\mathcal{A}}=\overbrace{\begin{bmatrix}
\hat{\mathcal{A}}/a &* &\ldots&*\\
* &*&\ldots&*\\
\vdots&\vdots&\ddots&\vdots\\
*&*&\ldots&*
\end{bmatrix}}^{2^m \text{matrix elements}},
\end{equation}
which corresponds to a dilation of the the $2^n\times 2^n$ matrix $\hat{\mathcal{A}}$ into a larger, $2^{n+m}\times2^{n+m}$ unitary matrix $\hat{U}_{\mathcal{A}}$ with $\norm*{\hat{\mathcal{A}}/a}\leq1$. 
The $*$ elements in Equation~\eqref{block encoded} represent proper sub-matrices for the $\hat{U}_{\mathcal{A}}$ matrix to be unitary and are of no direct relevance to our calculations. Additionally, $a$ is an appropriate renormalization factor of the matrix $\hat{\mathcal{A}}$ in the formation of the dilated unitary matrix $\hat{U}_{\mathcal{A}}$. Throughout the paper we will use the spectral norm of an operator, i.e its largest singular value, $\norm*{\hat{\mathcal{A}}}=max\{\sigma_{\mathcal{A}}\}$. 
Then, the application of $\hat{U}_{\mathcal{A}}$ on the composite state $\ket{0^{\otimes m}}\ket{\bol\psi}$, where $\ket{\bol\psi}$  is an $n$-qubit quantum state,  has the orthogonal decomposition
\begin{equation}\label{action of block encoded}
\hat{U}_{\mathcal{A}}\ket{0^{\otimes m}}\ket{\bol\psi}=\frac{1}{a}\ket{0^{\otimes m}}\hat{\mathcal{A}}\ket{\bol\psi}+\ket{\perp},
\end{equation}
where $\braket{\perp}{0^{\otimes m}}=0$. As with the Hadamard implementation in Figures~\ref{Hadamard product circuit} and ~\ref{multi Hadamard product circuit}, a successful projective measurement $\hat{P}=\ket{0^{\otimes m}}\bra{0^{\otimes m}}\otimes \hat{1}_{2^n\times2^n}$, in the first register with respect to the $0$-bit results in the normalized state $\frac{\hat{\mathcal{A}}\ket{\bol\psi}}{\norm{\hat{\mathcal{A}}\ket{\bol\psi}}}$ with success probability $p_{success}=\frac{\norm{\hat{\mathcal{A}}\ket{\bol\psi}}^2}{a^2}$. The respective quantum circuit for this post-selective process is depicted in Figure~\ref{block encoding circuit}.

Out of the various available  dilation techniques \cite{Schlimgen_2021,Schlimgen_2022,Jin_2023,Koukoutsis_2024,Koukoutsis_2025} for $\hat{U}_{\mathcal{A}}$, we employ an $m=1$ block encoding constructed via classical Singular Value Decomposition (SVD), combined with the LCU method \citep{Schlimgen_2022}. This selection is motivated because the non-trivial sub-matrices $\hat{A}_1$ and  $\hat{A}_2$ in Equations~\eqref{A1matrix} and ~\eqref{A2matrix} respectively are of small finite dimension. That leads to a classical implementation complexity of $\textit{O}(8^3)$ and $\textit{O}(10^3)$ respectively for the classical SVD. Therefore, the SVD decomposition of $\hat{\mathcal{A}}_1$ and  $\hat{\mathcal{A}}_2$ in Equations~\eqref{A1matrix} and ~\eqref{A2matrix} can be performed efficiently using a classical computer as part of a preprocessing routine.

Denoting the SVD decomposition of the matrix $\hat{\mathcal{A}}$ as $\hat{\mathcal{A}}=\hat{V^\dagger}\hat{\Sigma}\hat{W}$ where $\hat{V}$ and $\hat{W}$ are unitary matrices and $\hat{\Sigma}=diag(\bol\sigma)$ is a diagonal matrix containing the non-negative singular values ${\sigma}_i\geq0$, it can then be split into the sum of two unitary terms,
\begin{equation}\label{SVD spliting}
\hat{\mathcal{A}}=\frac{a}{2}\hat{V}^\dagger\hat{\Sigma}_+\hat{W}+\frac{a}{2}\hat{V}^\dagger\hat{\Sigma}_-\hat{W}, 
\end{equation}
where $a=\norm{\hat{\mathcal{A}}}$, and,
\begin{equation}\label{SVD angles}
\hat{\Sigma}_{\pm}=diag(e^{\pm i\bol\theta}), \quad \bol\theta=\arccos{{\Bar{\bol\sigma}}},\quad 0<\Bar{\sigma}_i\leq1.
\end{equation}

\begin{figure}[H]
\centering
\includegraphics[width=0.5\linewidth]{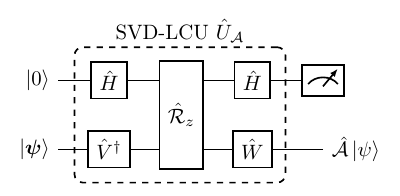}
    \caption{Explicit quantum circuit implementation for the SVD-LCU block encoding $\hat{U}_\mathcal{A}$ of Equation~\eqref{SVD-LCU block encoding}. }
    \label{SVD-LCU circuit}
\end{figure}

By applying the LCU technique~\citep{Childs_2012} to implement the two-term unitary sum in Equation~\eqref{SVD spliting}, only one extra qubit ($m=1$) is required to provide a $(1,\norm*{\hat{\mathcal{A}}})$-block encoding $\hat{U}_{\mathcal{A}}$,
\begin{equation}\label{SVD-LCU block encoding}
\hat{U}_{\mathcal{A}}=(\hat{H}\otimes\hat{V}^\dagger)\hat{U}_{select}(\hat{H}\otimes\hat{W}), 
\end{equation}
with $\hat{U}_{prep}^{LCU}=\hat{H}$ is the Hadamard gate and $\hat{U}^{LCU}_{select}$ is a diagonal matrix representing nested two-level $z$-rotations,
\begin{equation}\label{eq9}
\hat{U}_{select}=\ket{0}\bra{0}\otimes\hat{\Sigma}_+ +\ket{1}\bra{1}\otimes\hat{\Sigma}_{-}=\hat{\mathcal{R}}_z.
\end{equation}
As a result, the explicit quantum decomposition for the SVD-LCU block encoding $\hat{U}_\mathcal{A}$
is presented in Figure~\ref{SVD-LCU circuit}.

For an arbitrary matrix $\hat{\mathcal{A}}$, the decomposition cost of the SVD-LCU dilation within single qubit gates and CNOTs is $\textit{O}[(n+1)^2 2^{2n+1}]$~\citep{Schlimgen_2022}. However, the Lorenz system is finite-dimensional, requiring $3$ and $4$ qubits to encode the first and second order discretization schemes in Equations~\eqref{first order augmented},~\eqref{second order augmented}, respectively. Thus, the seemingly prohibitive exponential scaling becomes a constant, resulting into $\textit{O}(2^{11})$ and  $\textit{O}(5^2 2^9)$ elementary gates, respectively.

\subsection{Quantum circuit implementation of the second order time-marching scheme for the Lorenz equations}\label{sec:3.3}
\begin{figure}[H]
\centering
    \includegraphics[]{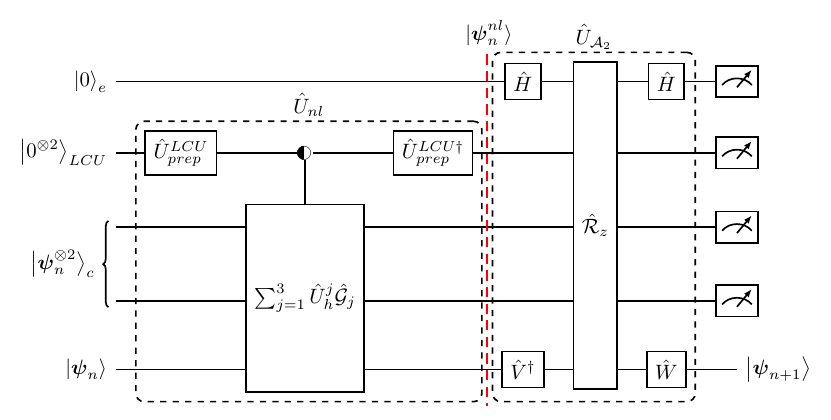}
    \caption{Quantum circuit implementation for the two-step evolution of Equation~\eqref{one step evolution block encoding}, regarding a single time step  evolution for the second-order discretized Lorenz model. The target register is at the bottom of the circuit, where the $\ket{\bol\psi_n}\to\ket{\bol\psi_{n+1}}$ advancement is achieved. }
    \label{U_1 circuit}
\end{figure}

Having set up all the essential elements,  the single-step quantum evolution  $\ket{\bol\psi_n}\to\ket{\bol\psi_{n+1}}$ as outlined in the Table~\ref{single step algorithm}, for the second order discretization scheme of the Lorenz model (Equations~\eqref{second order augmented}-~\eqref{c9-c10}) is given by,
\begin{equation}\label{one step evolution block encoding}
\hat{U}_{\mathcal{A}_2}\hat{U}_{nl}\ket{0}_e\ket*{0^{\otimes 2}}_{LCU}\ket*{\bol\psi_n^{\otimes 2}}_c\ket{\bol\psi_n}=\ket{0}_e\ket*{0^{\otimes 2}}_{LCU}\ket{0000}^{\otimes 2}_c\ket{\bol\psi_{n+1}}+\ket{\perp}_{e,LCU,c},
\end{equation}
with $\prescript{}{e,LCU,c}{\bra{\perp}}(\ket{0}_e\ket*{0^{\otimes 2}}_{LCU}\ket{0000}^{\otimes 2}_c)=0$. We can break down the various elements involved in the single-step evolution Equation~\eqref{one step evolution block encoding} as follows:
\begin{itemize}
    \item The second-order Lorenz system, given in Equations~\eqref{second order augmented}-~\eqref{c9-c10}, can be encoded using $4$ qubits analogous to the amplitude encoding representation in Equation~\eqref{Dirac notation}. Thus, any state $\ket{\bol\psi_n}$ is a $4$-qubit quantum state.
\item The nonlinear state $\ket*{\bol\psi^{nl}_n}$ in Equation~\eqref{state definition 2} consists of polynomial terms of third degree, in terms of the amplitudes $\mathbf{x}_n$. Subsequently, as discussed in Section~\ref{sec:3.1.2}, implementing $\ket*{\bol\psi^{nl}_n}$ requires a sum of Hadamard products between three copies of  $\ket{\bol\psi_n}$ interleaved with amplitude alternating $\hat{\mathcal{G}}$ gates,
\begin{equation}\label{general nonlinear opeator of Lorenz}
\hat{U}_{nl}=\sum_{j=1}^3\hat{U}_h^j\hat{\mathcal{G}}_j.
\end{equation}
The $\hat{\mathcal{G}}_j$ operators are three-fold gates acting on the $12$-qubit composite space of the $\ket{\bol\psi^{\otimes 3}}$ states, with $\hat{\mathcal{G}}_j=\hat{G}_{2j}\otimes\hat{G}_{1j}\otimes\hat{G}_{0j}$.
\item To implement $\hat{U}_{nl}$ we define a copy register $(c)$ that stores the two additional copied states $\ket*{\bol\psi_n^{\otimes 2}}_c$ where the $\hat{U}^j_h$ gates act and an LCU register $(LCU)$ to sum the respective terms, as presented in Figures~\ref{non linear operator circuit} and~\ref{explicit nonlinear state preparation circuit}. The resulting evolution is,
\begin{equation}\label{action of nonlinear operator for Lorenz}
\hat{U}_{nl}\ket*{0^{\otimes 2}}_{LCU}\ket*{\bol\psi_n^{\otimes 2}}_c\ket{\bol\psi_n}=\ket*{0^{\otimes 2}}_{LCU}\ket{0000}^{\otimes 2}_c\ket*{\bol\psi^{nl}}+\ket{\Tilde{\perp}}_{LCU,c}.
\end{equation}
\item Finally, by augmenting the nonlinear preparation process in Equation~\eqref{action of nonlinear operator for Lorenz} with the single qubit evolution register $(e)$, and applying the two-level variant of the block-encoding matrix $\hat{U}_{\mathcal{A}_2}$ from Equation~\eqref{SVD-LCU block encoding} that acts between the $(e)$ and the target register, we obtain,
\end{itemize}
\begin{equation}
\hat{U}_{\mathcal{A}_2}(\ket{0}_e\ket*{0^{\otimes 2}}_{LCU}\ket{0000}^{\otimes 2}_c\ket*{\bol\psi^{nl}}+\ket{0}_e\ket{\Tilde{\perp}})=\ket{0}_e\ket*{0^{\otimes 2}}_{LCU}\ket{0000}^{\otimes 2}_c\ket*{\bol\psi_{n+1}}+\ket{\perp}_{e,LCU,c}.
\end{equation}
Figure~\ref{U_1 circuit} illustrates the quantum circuit implementation corresponding to the aforementioned tasks for the second-order, single-step, time advancement of the Lorenz dynamics.

Lets now examine how we can extend the single time-step evolution into a complete time-marching quantum solver. Evidently, in order to implement the next consecutive step $\ket{\bol\psi_{n+1}}\to\ket{\bol\psi_{n+2}}$ (all states are considered normalized after a successful measurement) it is required to go though the $\hat{U}_{nl}$ that will prepare the intermediate nonlinear state $\ket*{\bol\psi^{nl}_{n+1}}$,
\begin{equation}\label{no-cloning}
\hat{U}_c\ket{0000}^{\otimes 2}_c\ket{\bol\psi_{n+1}}=\ket{\bol\psi_{n+1}}_c^{\otimes 2}\ket{\bol\psi_{n+1}}\xrightarrow{\hat{U}_{nl}}\ket{0000}^{\otimes 2}_c\ket*{\bol\psi^{nl}_{n+1}}
\end{equation}
Unfortunately, due to the no-cloning theorem in quantum computing~\citep{Nielsen_2010}, no such operator $\hat{U}_c$ exists that can satisfy the equality in Equation~\eqref{no-cloning} because the state $\ket{\bol\psi_{n+1}}$ is unknown. 

Subsequently, that means that  in order to perform the nonlinear transformation $\ket{\bol \psi_{n+1}}\to\ket*{\bol\psi^{nl}_{n+1}}$ many copies of the initial state $\ket{\bol\psi_n}$ have to be realized and evolved in a parallel way in order to serve as stepping stones for the nonlinear transformation. Therefore, based on the single step implementation $n\to n+1$ with $\hat{U}_1=\hat{U}_{\mathcal{A}_2}\hat{U}_{nl}$, the evolution operator $\hat{U}_j$ that defines the $j$-step advancement, $\hat{U}_j\ket{\bol\psi_n}\to\ket*{\bol\psi_{n+j}}$ can only be defined recursively,
\begin{equation}\label{recursive U_j}
\hat{U}_{j}=\hat{U}_1\hat{U}_{j-1}^{\otimes 3}.
\end{equation}
Indeed,
\begin{equation}
\hat{U}_j\ket{0}_e\ket*{0^{\otimes 2}}_{LCU}\ket*{\bol\psi_n^{\otimes 2}}_c\ket{\bol\psi_n}\to\hat{U}_1\ket{0}_e\ket*{0^{\otimes 2}}_{LCU}\ket*{\bol\psi_{n-j-1}}^{\otimes 2}_c\ket*{\bol\psi_{n-j-1}}\to\ket*{\bol\psi_{n+j}}.
\end{equation}

\begin{figure}[H]
\centering
\includegraphics[width=\linewidth]{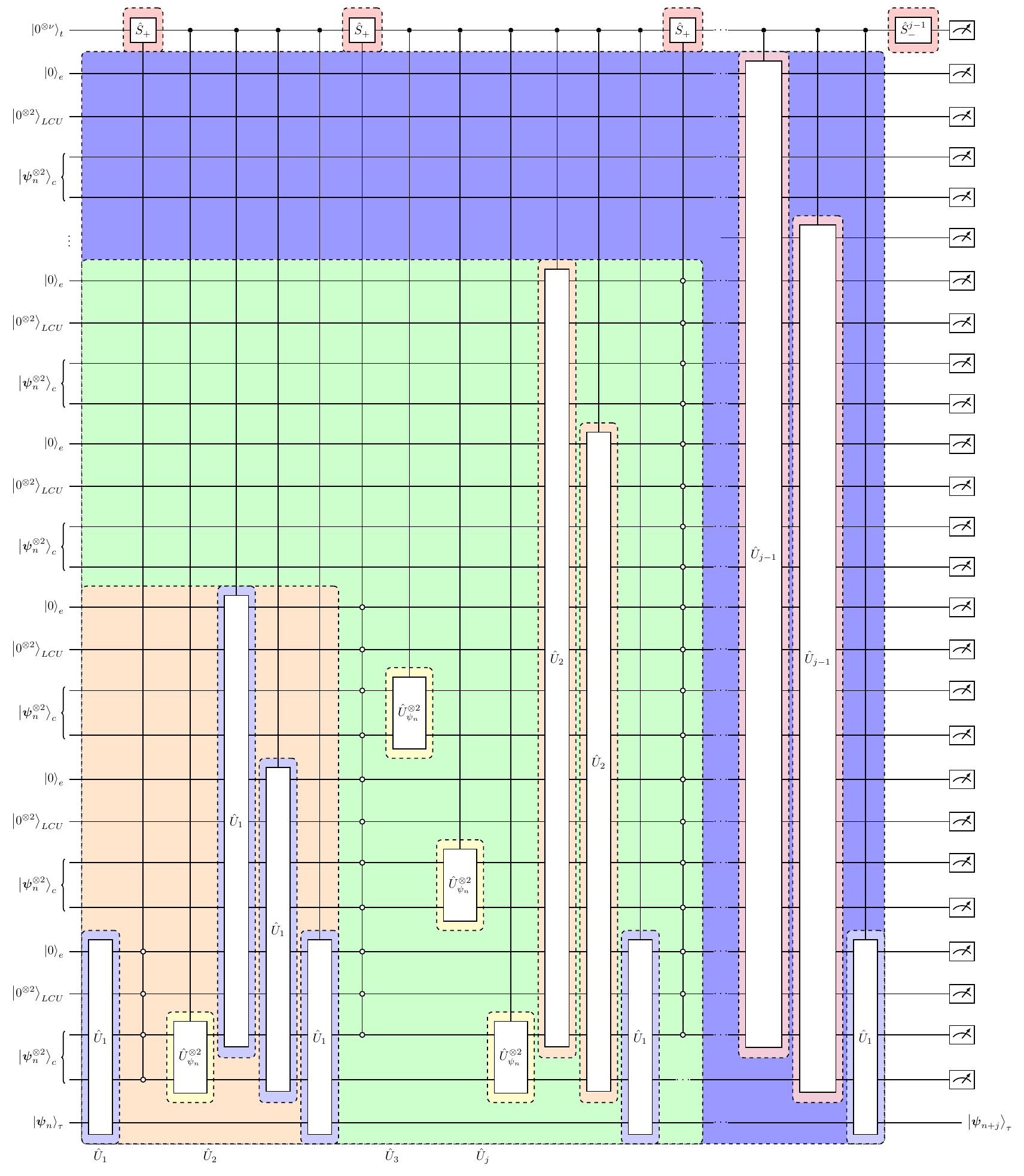}
    \caption{Schematic quantum circuit implementation of the time-marching quantum algorithm for the steps described in $1$-$7$ in Table~\ref{full algorithm}. The explicit implementation of the single time step evolution operator $\hat{U}_1=\hat{U}_{\mathcal{A}_2}\hat{U}_{nl}$ has been presented in Figure~\ref{U_1 circuit}. The coloring reveals the recursive nested structure. For example the operator $\hat{U}_3$ (green color shade) contains $2$ applications of the $\hat{U}_2$ gates and one of $\hat{U}_1$. In turn the $\hat{U}_2$ operators (orange color shade) contain $4$ $\hat{U}_1$ gates. The shift $\hat{S}$ gates acting on the clock $(t)$ register are colored in red and the preparation operators $\hat{U}_{\psi_n}^{\otimes 2}$ in yellow.}
    \label{total evolution cicuit}
\end{figure}

Taking into consideration the no-cloning theorem and Equation~\eqref{recursive U_j}, a quantum time-marching algorithm is delineated in Table~\ref{full algorithm}.
\begin{table}[H] 
\caption{Quantum algorithm for the simulation of the second order discretized Lorenz system~\eqref{second order augmented}. \label{full algorithm}}.
\begin{tabularx}{\textwidth}{L}
\toprule
\textbf{Time-marching evolution $n\to n+j$.}	\\
\midrule
$1$. Initial target state $\leftarrow$ $\ket{\bol \psi_n}_{\tau}$.\\
$2$. Prepare a clock-time $(t)$ register $\leftarrow$ $\ket{0^{\nu}}_t$ with $\nu=[\log_2{j}]+1$ and
 $2j-1$ copies of the $(e)$, $(LCU)$ and $(c)$ registers $\leftarrow$ $(\ket{0}_e\ket*{0^{\otimes 2}}_{LCU}\ket*{\bol\psi_n^{\otimes 2}}_c\ket{\bol\psi_n})^{\otimes {2j-1}}$. The target state is chosen to be the least significant. \\
$3$. Evolve the target state by acting with the single-step evolution operator $\hat{U}_1$ using the least significant of the $(e)$,$(LCU)$,$(c)$-registers as dictated in Equation~\eqref{one step evolution block encoding}, $\ket{\bol\psi_n}\to\ket{\bol\psi_{n+1}}$.\\
$4$. Update the clock $(t)$ register, $C\hat{S}^+\ket{0}_t\to\ket{1}_t$ to keep track of the time advancement of the target state controlled by the $0$-bits in the previously used least significant, $i=1$ of the $(e)$,$(LCU)$,$(c)$-registers.\\
$5$. Replenish the states $\ket{\bol\psi_n}^{\otimes 2}_c$ in the  $(c)$ register with a control operator $C\hat{U}_{\psi_n}^{\otimes 2}$ with respect to the $1$-bit state in the $(t)$ register with $\hat{U}_{\psi_n}\ket{0000}=\ket{\bol\psi_n}$.\\
$6$. Recursively repeat $j-1$ times the steps $3$, $4$ and $5$, using $\hat{U}_j$ in Equation~\eqref{recursive U_j} and $2j-1$ ancillary registers for each iteration.\\
$7$. Initialize the clock $(t)$ register to the $\ket{0}_t$ state and measure all registers in respect to the $0$-bit states to obtain, $\ket{\bol\psi_n}\to\ket*{\bol\psi_{n+j}}$.\\
\bottomrule
\end{tabularx}
\end{table}
The respective quantum circuit implementation for the time-marching algorithm in Table~\ref{full algorithm} is presented in Figure~\ref{total evolution cicuit}. A characteristic feature of the algorithm is that the copies of the initial state  $\ket{\bol\psi^{\otimes 2}_n}$ in the $(c)$ registers, are not discarded after each time step evolution. Instead, they are re-prepared using the $C\hat{U}_{\psi_n}^{\otimes 2}$ operation such that the $(c)$ registers contain a sufficient number of $\ket{\bol\psi_n}$ states to be used for creating the $\ket*{\bol\psi_{n+j-1}}^{\otimes 2}_c$ in the least significant of the $(c)$ registers. Then, applying $\hat{U}_1$  on the $\ket*{\bol\psi_{n+j-1}}^{\otimes 2}_c\ket*{\bol\psi_{n+j-1}}_{\tau}$ state we obtain the $j$-time update. We build upon this logic by schematically presenting the re-usage of the copy register for a two-step, $n\to n+2$, advancement following Figure~\ref{total evolution cicuit}:
\begin{align}
&\hdots\ket{\bol\psi_n}^{\otimes 2}_c\hdots\ket{\bol\psi_n}^{\otimes 2}_c\underbrace{\ket{\bol\psi_n}_{\tau}}_{\text{target state}}\\
\xrightarrow{\hat{U}_1}&\hdots\ket{\bol\psi_n}^{\otimes 2}_c\hdots\underbrace{\ket{0000}_c^{\otimes 2}}_{\text{re-use}}\ket{\bol\psi_{n+1}}_{\tau}+\hdots\label{quantum algorithm step 1}\\
\xrightarrow{C\hat{U}_{\psi_n}^{\otimes 2}}&\hdots\ket{\bol\psi_n}_c\underbrace{\ket{\bol\psi_n}_c\hdots\ket{\bol\psi_n}_c}_{\text{available copies}}\underbrace{\ket{\bol\psi_n}_c}_{\text{intermediate target state}}\ket{\bol\psi_{n+1}}_{\tau}+\hdots\\
\xrightarrow{\hat{U}_1}&\hdots\ket{\bol\psi_n}_c\underbrace{\ket{0000}\hdots\ket{0000}}_{\text{re-use}}\ket{\bol\psi_{n+1}}_c\ket{\bol\psi_{n+1}}_{\tau}+\hdots\\
\xrightarrow{C\hat{U}_{\psi_n}^{\otimes 2}}&\hdots\underbrace{\ket{\bol\psi_n}^{\otimes 2}_c}_{\text{available copies}}\hdots\underbrace{\ket{\bol\psi_n}_c}_{\text{intermediate target state}}\ket{\bol\psi_{n+1}}_c\ket{\bol\psi_{n+1}}_{\tau}+\hdots\\
\xrightarrow{\hat{U}_1}& \hdots\ket{0000}^{\otimes 2}_c\hdots\underbrace{\ket{\bol\psi_{n+1}}_c^{\otimes 2}\ket{\bol\psi_{n+1}}_{\tau}}_{\text{evolution to $n+2$}}+\hdots\\
\xrightarrow{\hat{U}_1}&\hdots\ket{0000}^{\otimes 2}_c\hdots\ket{0000}_c^{\otimes 2}\ket{\bol\psi_{n+2}}_{\tau}+\hdots\label{quantum algorithm step end}
\end{align}
This re-usage of qubits~\citep{Esmaeilifar_2024} allows for parallel processing of the quantum states and optimal scaling in the numbers of required copies of the state $\ket{\bol\psi_n}$, as will be discussed in the next section.

\subsection{Complexity considerations and discussion}\label{sec:3.4}
In quantum algorithmic architectures incorporating many copies of the initial state $\ket{\bol\psi_n}$, the important aspects dictating the efficiency of the algorithm  are the number of copies that the algorithm requires (circuit width) and the number of elementary operations (circuit depth).

Due to the re-usage of the states in the $(c)$ registers, as presented in Equations~\eqref{quantum algorithm step 1}-~\eqref{quantum algorithm step end}, the required number of copies, $\mathcal{N}_{copies}$, of the $4$-qubit state $\ket{\bol\psi_n}$ scales linearly with respect to the number of time integration steps $N_t$ for a simulation $n\to n+N_t$,
\begin{equation}\label{number of copies}
\mathcal{N}_{copies}(N_t)=4N_t-1.
\end{equation}
This linear dependence of the number of copies on $N_t$ establishes an exponential improvement over the results of Ref.~\citep{Leyton_2008} and a quadratic improvement compared to Refs.~\citep{Lloyd_2020,Andress_2025}. This technique was firstly demonstrated in Ref.~\citep{Esmaeilifar_2024}. Taking into consideration the ancillary qubits in the $(t)$ $(LCU)$ and $(e)$ registers, the total number of working quantum resources is,
\begin{equation}\label{total number of qubits}
\mathcal{N}_{qubits}(N_t)=(2N_t-1)(\underbrace{8}_{\text{$(c)$}}+\underbrace{1}_{\text{$(e)$}}+\underbrace{2}_{\text{$(LCU)$}})+\underbrace{\log_2N_t+1}_{\text{$(t)$}}+\underbrace{4}_{\text{target}}=22N_t+\log_2N_t-6.
\end{equation}
Therefore, the total number of operational qubits for the overall quantum time-integration scheme  scales as $\mathcal{N}_{qubits}(N_t)=\textit{O}(N_t)$.

On the other hand, the recursive structure of the time-marching algorithm in Table~\ref{full algorithm} enables a direct count of the operators $\hat{U}_1$, $C\hat{U}_{\psi_n}^{\otimes 2}$ and $\hat{S}$ which are highlighted in blue, yellow and red respectively in Figure~\ref{total evolution cicuit}. Specifically, the number $\mathcal{N}_{U_1}$ of the $\hat{U}_1$ gates that comprise each $\hat{U}_j$ is,
\begin{equation}\label{number U1 for j step}
\mathcal{N}_{U_1} (j)=3^{j-1}.
\end{equation}
Thus, the total number of $\hat{U}_1$ gates for $N_t$ time-steps reads,
\begin{equation}\label{U_1 scaling}
\mathcal{N}_{U_1}(N_t)=\sum_{j=1}^{N_t}3^{j-1}=\frac{3^{N_t}}{2}.
\end{equation}
Accordingly, the number of $\hat{S}$ gates (in red in Figure~\ref{total evolution cicuit}) acting on the clock $(t)$ register is $\mathcal{N}_S(N_t)=2N_t$ and, similarly with Equation~\eqref{U_1 scaling} due to recurrence, the number of the $\hat{U}_{\psi_n}^{\otimes 2}$ gates (in yellow in Figure~\ref{total evolution cicuit}) scales as 
\begin{equation}
 \mathcal{N}_{U_{\psi_n}^{\otimes 2}}(N_t)=3^{N_t-1}-N_t.
\end{equation}
As a result, the total number of queries to the building blocks of our algorithm is,
\begin{equation}\label{complexity scaling}
\mathcal{N}_{queries}=\frac{5}{6}3^{N_t}+N_t=\textit{O}(3^{N_t}).
\end{equation}
Finally, since a second order time discetization scheme has been employed,  the total simulation error $\varepsilon$ in terms of the total simulation time $T$ is $\varepsilon=T^2/N_t$.

The decomposition cost of the $\hat{U}_{\delta t}$ and $C\hat{U}_{\psi_n}$ operators into elementary gates follows is constant because each of the $\hat{U}_{nl}$, $\hat{U}_{\mathcal{A}_2}$ and $\hat{U}_{\psi_n}$ operators act in the $4$-qubit space (see Sections~\ref{sec:3.1.1},~\ref{sec:3.1.2} and~\ref{sec:3.2}). The corresponding circuit implementation has been explicitly provided in Figures~\ref{non linear operator circuit} and ~\ref{SVD-LCU circuit},~\ref{U_1 circuit}. Thus, the resulting implementation scaling into CNOTs and single-qubit gates scales as $\textit{O}(28\cdot2^9N^2_t)=\textit{O}(60\cdot16^2)$. Implementation of the  $\hat{S}$ operator scales as $\textit{O}(\nu^2)=\textit{O}(\log^2_2N_t)$.

Consequently, the proposed quantum solver for the second-order discretized Lorenz system requires an operational circuit width that scales linearly with the time-integration steps and with overall complexity,
\begin{equation}\label{total complexity}
\textit{O}(60d^2p^{N_t}, T^2/\epsilon), \quad p=3,\quad d=16, \quad \mathcal{N}_{copies}(N_t)=4N_t-1.
\end{equation}
Notice that $d=16$ is the dimension of the second order discretized Lorenz system in Equation~\eqref{state definition 2}, and $p=3$ is the maximum  polynomial degree in the nonlinear state $\ket*{\bol\psi_n^{nl}}$.

The complexity scaling presented in Equation~\eqref{total complexity} is characteristic of the proposed quantum algorithmic scheme. Consider solving a polynomial nonlinear system of ODEs of dimension $d_s$ with a maximum degree $p_s$. Then, employing a $K$-th order discretization scheme the produced nonlinear state $\ket*{\bol\psi^{nl}}$ has polynomial degree $p=p_s+K$ and  dimension $d$. Therefore, the augmented system (similar to that of Equations~\eqref{first order augmented} and~\eqref{second order augmented}) can be represented within $n=\log_2 d$ qubits. To prepare the state $\ket*{\bol\psi^{nl}}$ through the Hadamard product we need $p$ copies. Therefore, the complexity of the general quantum solver in analogy with Equation~\eqref{total complexity} is given by,
\begin{equation}\label{general algoritmic complexity}
\textit{O}\Big[4^{n}n^2p^{N_t}, \Big(\frac{T^K}{\varepsilon}\Big)^\frac{1}{K-1}\Big]=\textit{O}\Big[d^2\log^2_2dp^{N_t}, \Big(\frac{T^K}{\varepsilon}\Big)^\frac{1}{K-1}\Big],\quad \mathcal{N}_{copies}(N_t)=\textit{O}(N_t). 
\end{equation}
When $d_s<<d$,  the scaling with respect the dimension of the original nonlinear system is $\textit{O}[polylog(d_s)p^{N_t}]$. As a result, the quantum solver exhibits an exponential speed-up  over classical ODE solvers, which typically scale exponentially with the dimension $d_s$,  while requiring only linear dependence on the number of state copies in respect to the integration steps $N_t$. To the best of our knowledge, the complexity scaling presented in Equation~\eqref{general algoritmic complexity} represents the most efficient result reported in the literature for quantum nonlinear solvers.

Using quantum compression gadgets~\citep{Low_2019,Fang_2023} it is anticipated that the above scalings in Equations~\eqref{number of copies} -~\eqref{general algoritmic complexity} can be further improved.

The final feature of the algorithm  to be discussed is its post-selective, and therefore, probabilistic nature. As illustrated in Figure~\ref{total evolution cicuit}, obtaining the correct result requires a single successful measurement at the output in all ancillary register in respect to the $0$-bits. While the 
continuum Lorenz nonlinear differential equations yield a contraction map in time, as shown in Equation~\eqref{exponential decay of the pahse space}, this does not hold for the discretized Lorenz system, i.e., for the operator 
$\hat{\mathcal{A}}_2$ in Equation~\eqref{A2matrix},
As a result, a normalization factor $a=\sigma_{max}$, where $\sigma_{max}$ is the maximum singular value of $\hat{\mathcal{A}}_2$, dictates the success probability of the SVD-LCU block encoding $\hat{U}_{\mathcal{A}_2}$, as described in Section~\ref{sec:3.2}. Since $\hat{U}_{\mathcal{A}_2}$ is a component in the fundamental algorithmic composite block $\hat{U}_1=\hat{U}_{\mathcal{A}_2}\hat{U}_{nl}$ in Equation~\eqref{one step evolution block encoding}, and the number of $\hat{U}_1$ repetitions scale according to Equation~\eqref{U_1 scaling}, then the overall success implementation probability is,
\begin{equation}\label{success probability}
p_{success}(N_t)\propto \Big(\frac{1}{\sigma^2_{max}}\Big)^{3^{N_t}}.
\end{equation}

\begin{figure}[H]
\centering
    \includegraphics[width=0.4\linewidth]{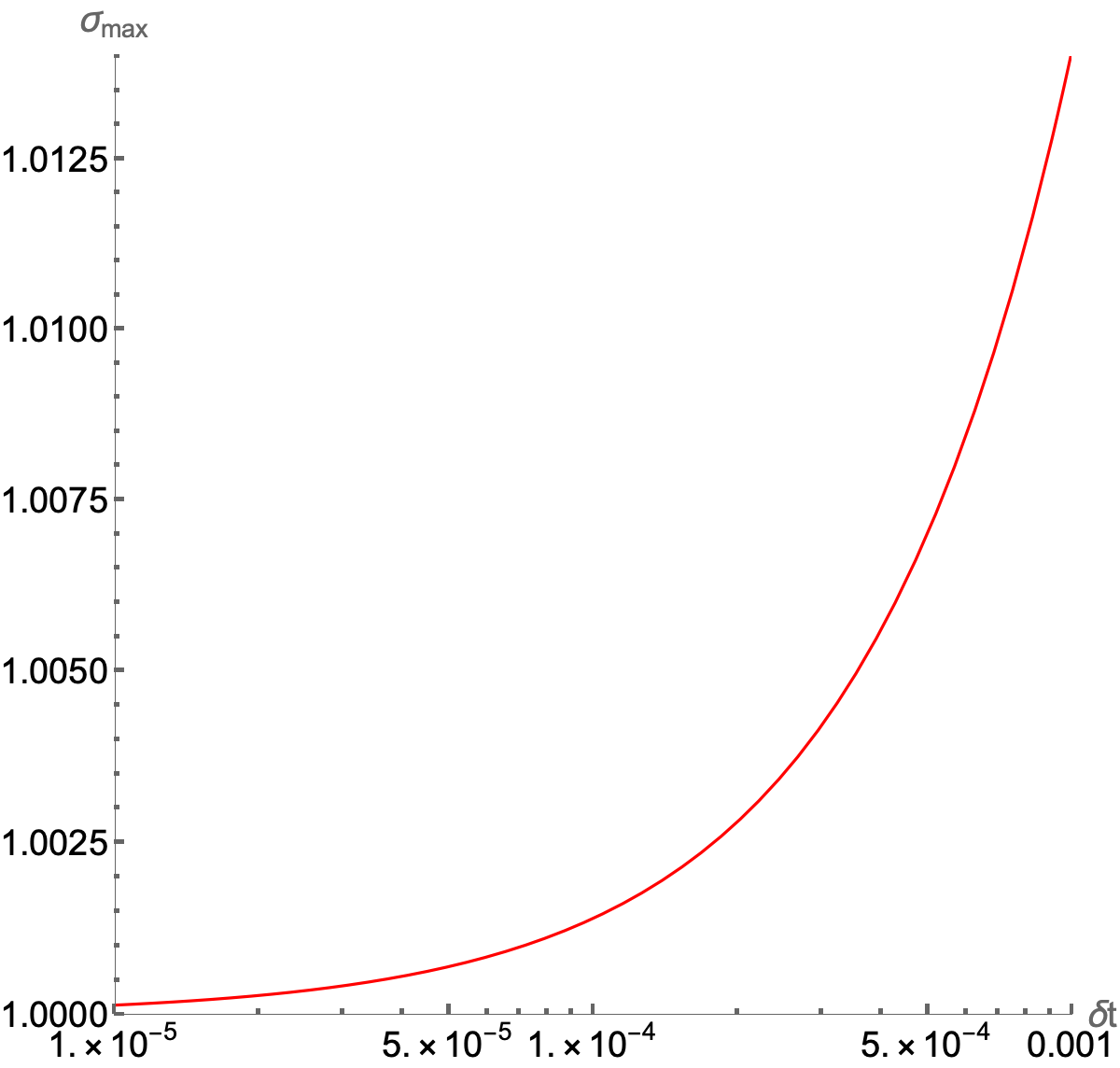}
    \caption{The dependence of the largest singular value of $\hat{\mathcal{A}}_2$ matrix in Equation~\eqref{A2matrix}, $\sigma_{max}$, on the discretization time step $\delta t$ for the typical time steps used in Section~\ref{sec:4}.}
    \label{max singular value}
\end{figure}

In turn, the value of $\sigma_{max}$, for the $\hat{\mathcal{A}_2}$ matrix depends on the chosen time discretization step $\delta t$  as dictated in Equations~\eqref{a1}-~\eqref{c9-c10}. The behavior of $\sigma_{max}(\delta t)$  as a function of the time step $\delta t$ is shown in Figure~\ref{max singular value}.  Selecting a time step $\delta t<<1$ improves the accuracy of the algorithm and results in a $\sigma_{max}<1.01$ according to Figure~\ref{max singular value}. Thus, the  implementation success probability in Equation~\eqref{success probability} is improved. However, choosing an extremely small $\delta t<<1$, simultaneously increases the number of iteration steps $N_t$ required for a total simulation time $t\to t+T,\,\,T=N_t\delta t$. However, due to the exponential dependence of the success probability on $N_t$ in Equation~\eqref{success probability}, this eventually leads to a significant reduction in the overall probability for a successful implementation of our quantum algorithm.

Consequently, at the output of the algorithm, amplitude amplification~\citep{Brassard_2002} of the state with all ancillary qubits in the $0$-bit  is required, in order for a successful measurement to occur with a non-vanishing probability.

\section{Numerical Demonstrations}\label{sec:4}
While the proposed quantum solver exhibits a quantum speed-up in terms of the dimension of the system compared to the classical ODE solvers, the exponential scaling in the number of operations with the integration steps in Equation~\eqref{total complexity} puts limitations to the quantum implementation of our algorithm in the Noisy Intermediate-Scale Quantum  (NISQ) era, with relatively high error rates and relatively small number of allowed operations. Therefore, we resort to classical implementation of the quantum algorithm presented in Section~\ref{sec:3.2} for the Lorenz second-order scheme, Equations~\eqref{second order augmented} -~\eqref{c9-c10}, for various values of $\beta$ at $\sigma = 10$ and $\rho = 28$.

\begin{figure}[H]
\centering
    \includegraphics[width=0.5\linewidth]{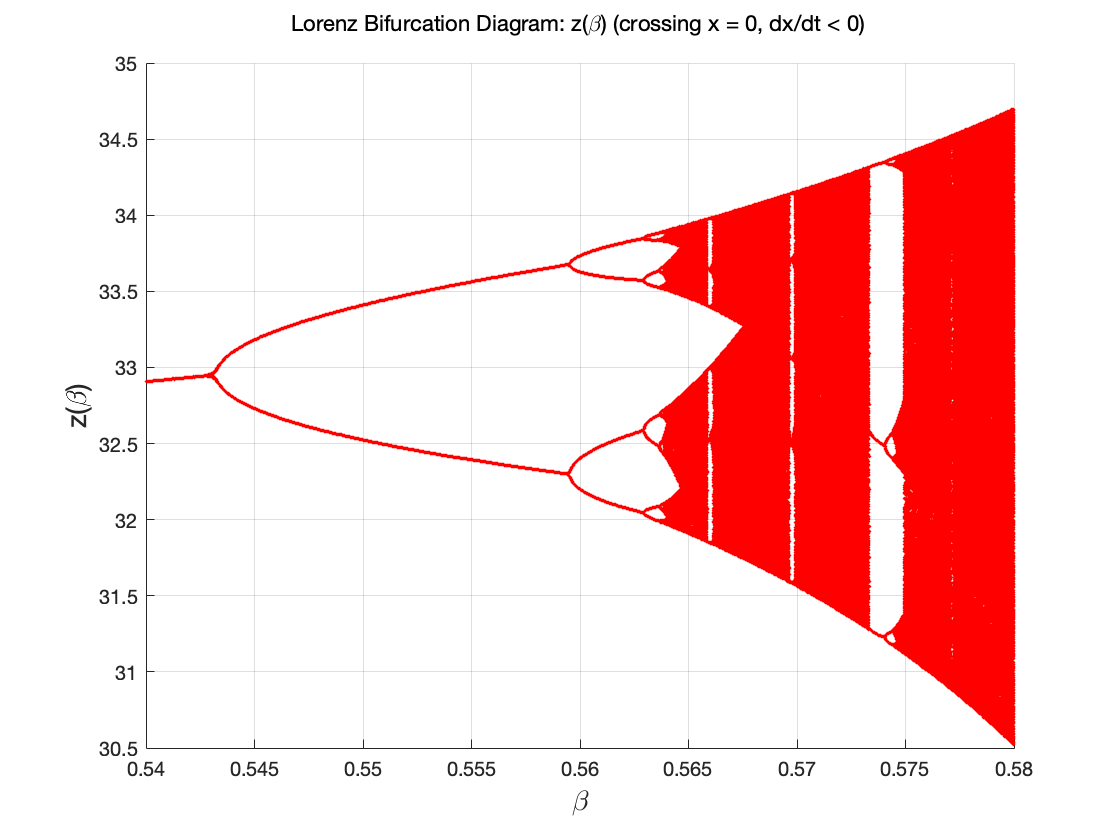}
    \caption{The bifurcation diagram for $z(\beta)$ for $0.54 < \beta < 0.58$, with $\sigma = 10$, $\rho = 28$.  The bifurcation diagrams for $x(\beta)$ and $y(\beta)$ are qualitatively similar and with exactly the same bifurcation points in $\beta$.}
    \label{birfucation diagram}
\end{figure}

Following the work of Moysis~\citep{Moysis_2025}, we plot the bifurcation diagram, Figure~\ref{birfucation diagram}, for $z(\beta)$ as a function of the parameter $\beta$ in the range $0.54 < \beta < 0.58$ with $\sigma = 10$ and $\rho=28$. This bifurcation diagram 
is a Poincare plot of the Lorenz attractor as its orbit intersects the plane $x  = 0$ with $dx/dt < 0$.  
One can readily identify  the bifurcation from period-1 $(P1)$ to period-2 $(P2)$ limit cycle for $\beta \approx  0.542$.  A bifurcation from
$P2$ - $P2^2$  around $\beta \approx 0.558$, and a $P2^2-P2^3$ bifurcation for $\beta \approx 0.5624$. Indeed, in this small parameter window
$0.54 < \beta < 0.58$,  this bifurcation diagram is formally similar to the well known period-doubling bifurcation route to chaos for the $1$D discrete logistic map $\eta_{n+1} = 4 \mu  \eta_n (1-\eta_n)$ as the parameter $\mu$ is varied.

In the following, we shall compare the Lorenz  regular (limit cycles) and chaotic attractors as determined from our second order fixed time scale solution to 
Equations~\eqref{second order augmented}-~\eqref{c9-c10}
with those derived  from the higher order adaptive time scale ODE solver in $Mathematica$. For values of $\beta$ giving rise to a chaotic attractor (e.g., $\beta = 0.58$), the $2$nd-order finite difference scheme will require a smaller time step $\delta t = 2.5 \cross 10^{-4}$ than is needed to determine the periodic limit cycles (where  $\delta t = 10^{-3}$). 

\subsection{Chaotic Attractor : $ \beta = 0.58 $}
\begin{figure}[H]
\centering
\subfloat[\centering]{\includegraphics[width=0.35\linewidth]{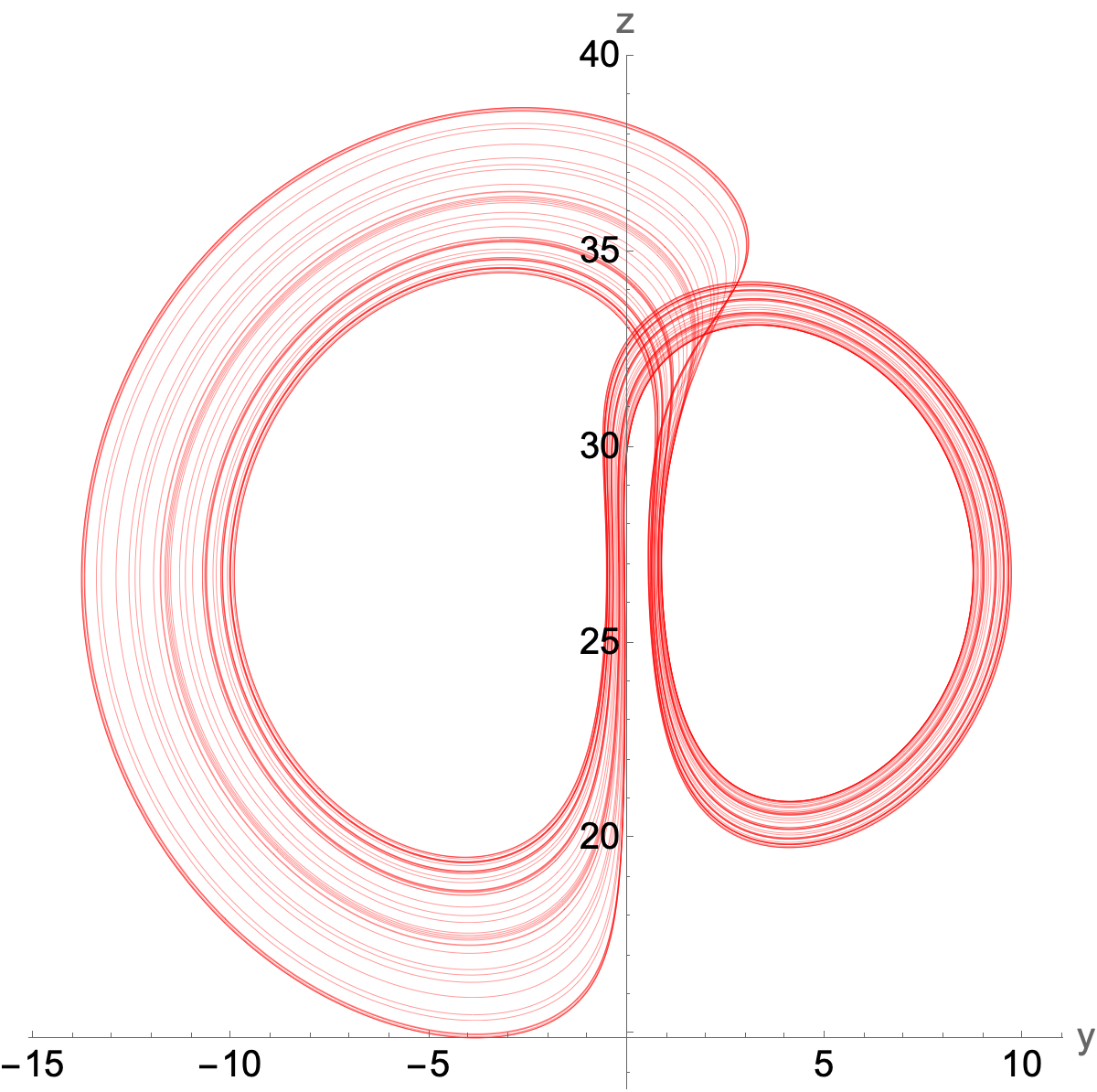}
  }
  \subfloat[\centering]{\includegraphics[width=0.32\linewidth]{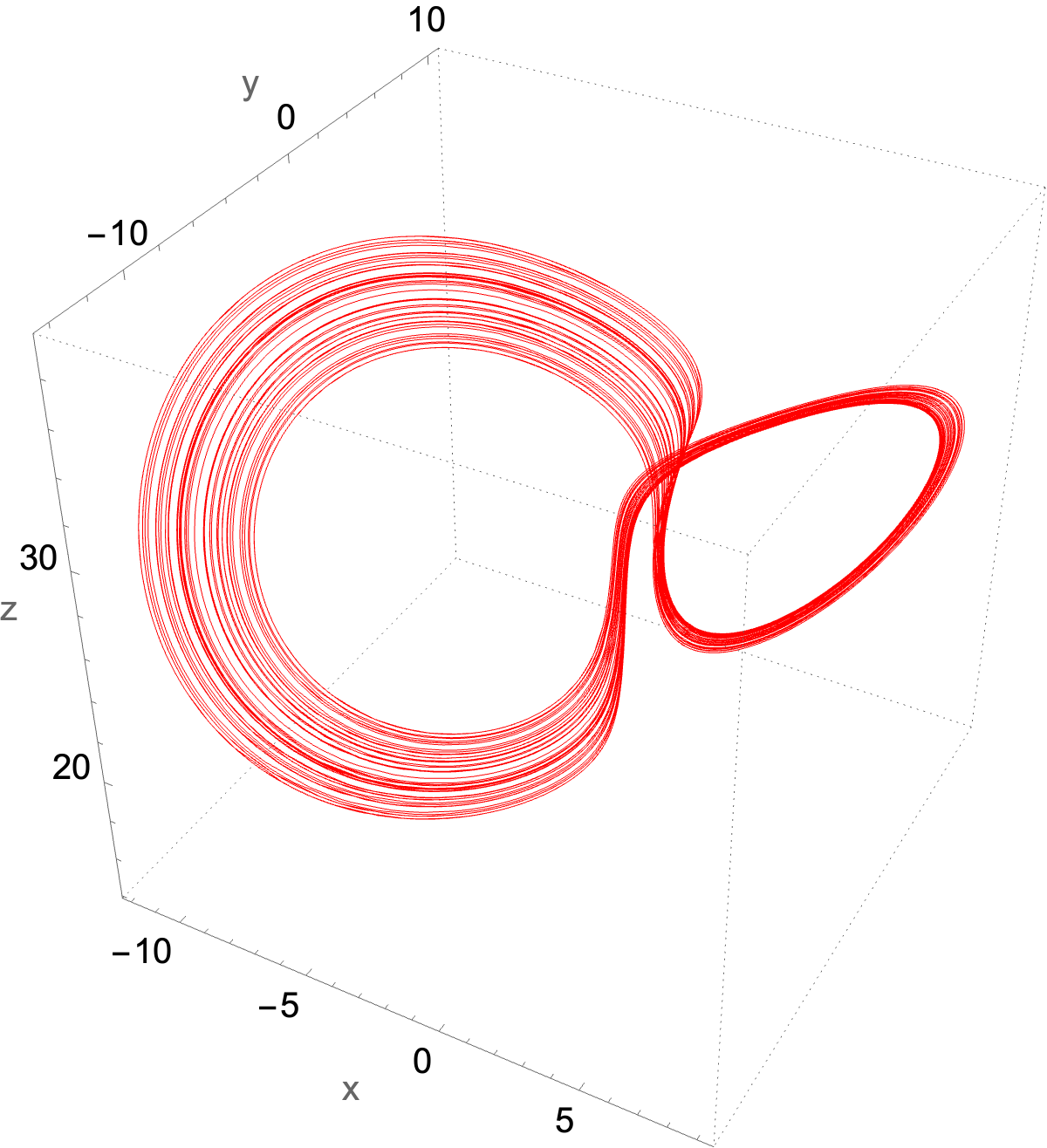}
  }
    \caption{The chaotic attractor for $\beta = 0.58$, $\sigma = 10$ and $\rho = 28$, showing both (\textbf{a}) its  y-z projection and (\textbf{b}) the $3$D plot.  Initial conditions: $x(0) = 0.1, y(0) = -1.1, z(0) = 10.1$.  These plots were generated by a higher order adaptive time $Mathematica$ ODE solver. The lobes of the attractor are somewhat anchored around the two unstable fixed points of the Lorenz equations $\left ( \pm \sqrt{\beta (\rho-1)}, \pm \sqrt{\beta (\rho-1)}, \rho-1 \right) $.}
    \label{Mathematica chaotic attractor}
\end{figure}

\begin{figure}[H]
\centering
\subfloat[\centering]{\includegraphics[width=0.35\linewidth]{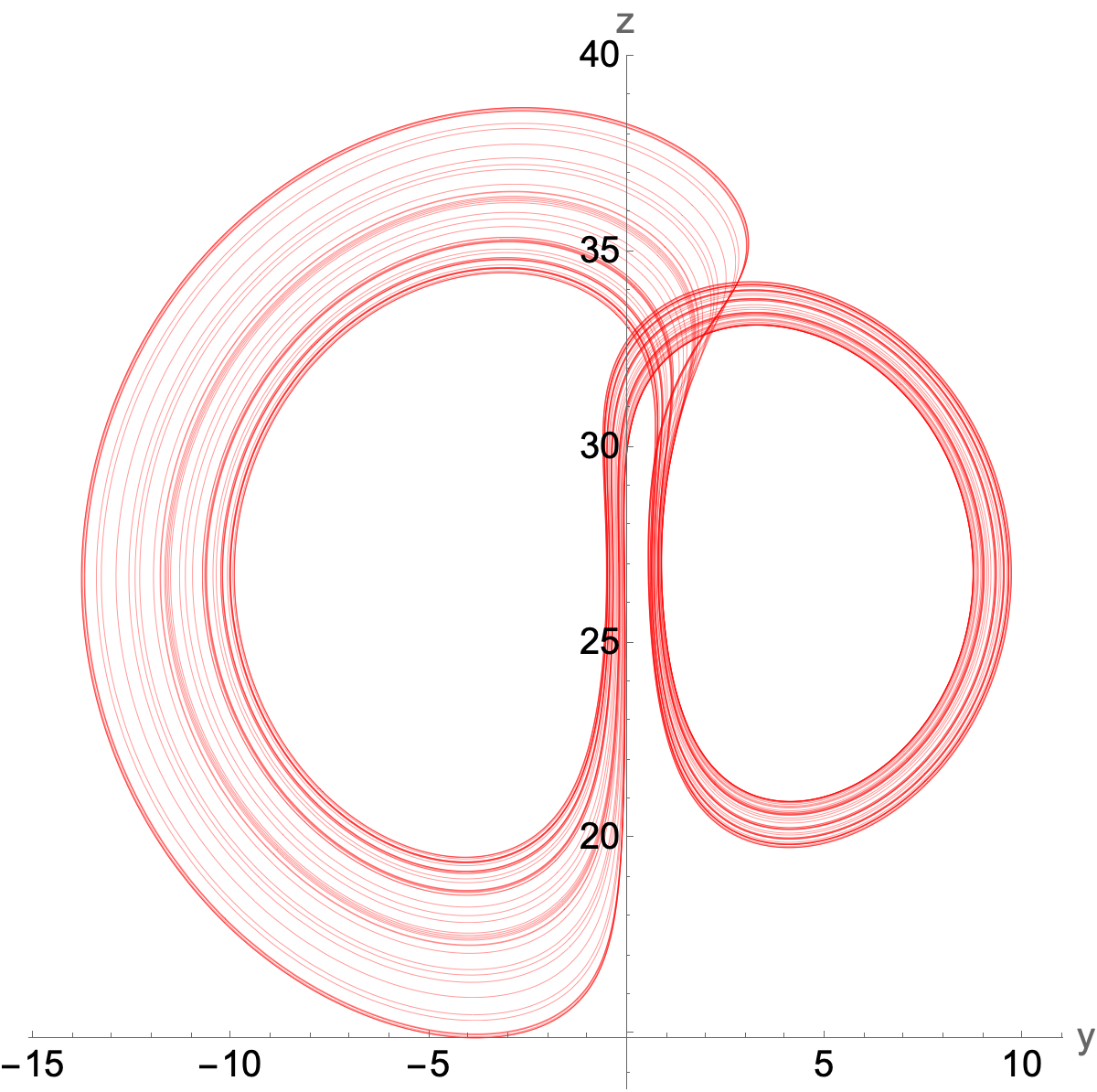}
  }
  \subfloat[\centering]{\includegraphics[width=0.32\linewidth]{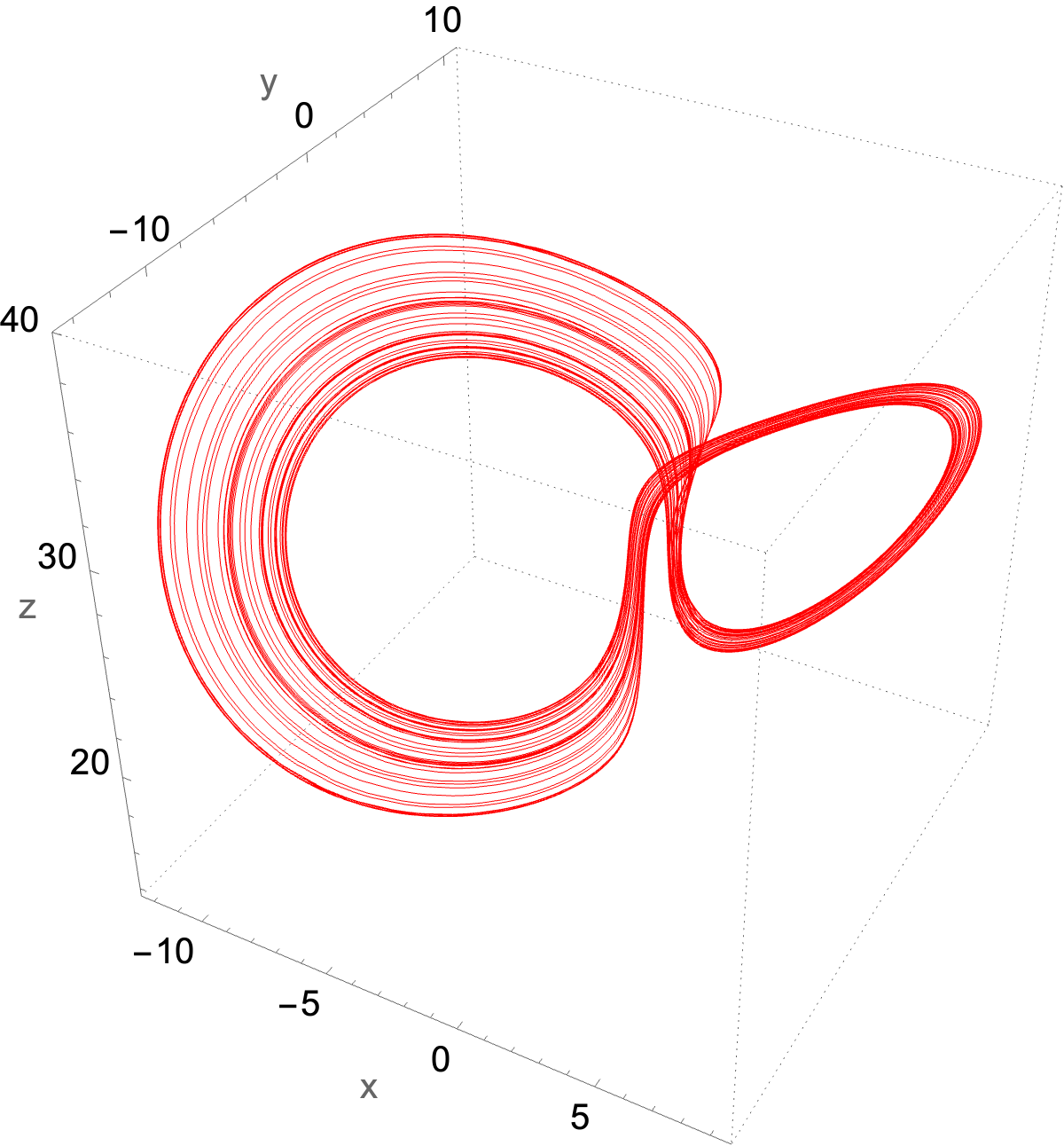}
  }
    \caption{The chaotic attractor for $\beta = 0.58$, and with the same initial conditions as in Figure~\ref{Mathematica chaotic attractor}. However, this attractor is determined from the classical implementation of the $2$nd order recursive quantum algorithm with fixed time step $\delta t = 2.5 \cross 10^{-4}$.}
    \label{2nd order chaotic attractor}
\end{figure}

The $\beta$-bifurcation diagram, Figure~\ref{birfucation diagram}, indicates that we should expect chaos for $\beta = 0.58$.  Indeed, we find such an attractor using the  higher order adaptive time ordinary differential equation (ODE) solver, as presented in Figure~\ref{Mathematica chaotic attractor}.  The asymmetry in the lobes make this attractor look considerably different than the standard picture of the ``butterfly" attractor for the more common parameter value $\beta = 8/3$. Here, both the $2$D phase plot in the $y-z$ plane as well as the full 3D attractor in the $x-y-z$ plane for initial conditions $x(0)=0.1, y(0)= -1.1, z(0) = 10.1$ are presented in Figure~\ref{Mathematica chaotic attractor}. Notice that, by changing just the initial $z$ to $z(0)=1.1$, but keeping $x(0) = 0.1$ and $y(0) = -1.1$, then the resulting Lorenz attractor is still chaotic, but with mirror symmetry.
Implementing the quantum solver for the second-order finite difference scheme, we find excellent qualitative agreement between the $2$nd-order generated chaotic attractor in Figure~\ref{2nd order chaotic attractor} and that generated by the adaptive grid higher order ODE solver used in Figure~\ref{Mathematica chaotic attractor}. 
This signifies that a successful quantum implementation of our method has the potential to resolve the chaotic region for the Lorenz system, preserving the basic structural characteristics of the chaotic attractor.
\subsection{Limit cycles}
Aside from the chaotic attractors, different structures that arises in the region of regular dynamics of the Lorenz system are limit cycles. Limit cycles are isolated periodic orbits, which also exhibit sensitivity to perturbations. For example, in Figure~\ref{birfucation diagram}, as the parameter $\beta$ increases there is a transition through bifurcations from the regular dynamics limit cycles to chaos.

In the following, some of the period-doubling non-chaotic Lorenz attractors are considered, with initial conditions $x(0)=0.1$, $y(0)= -1.1$, $z(0) = 1.1$, using the presented second-order scheme.
\subsubsection{P1 - Limit cycle : $\beta = 0.52$}
\begin{figure}[H]
\centering
\subfloat[\centering]{\includegraphics[width=0.35\linewidth]{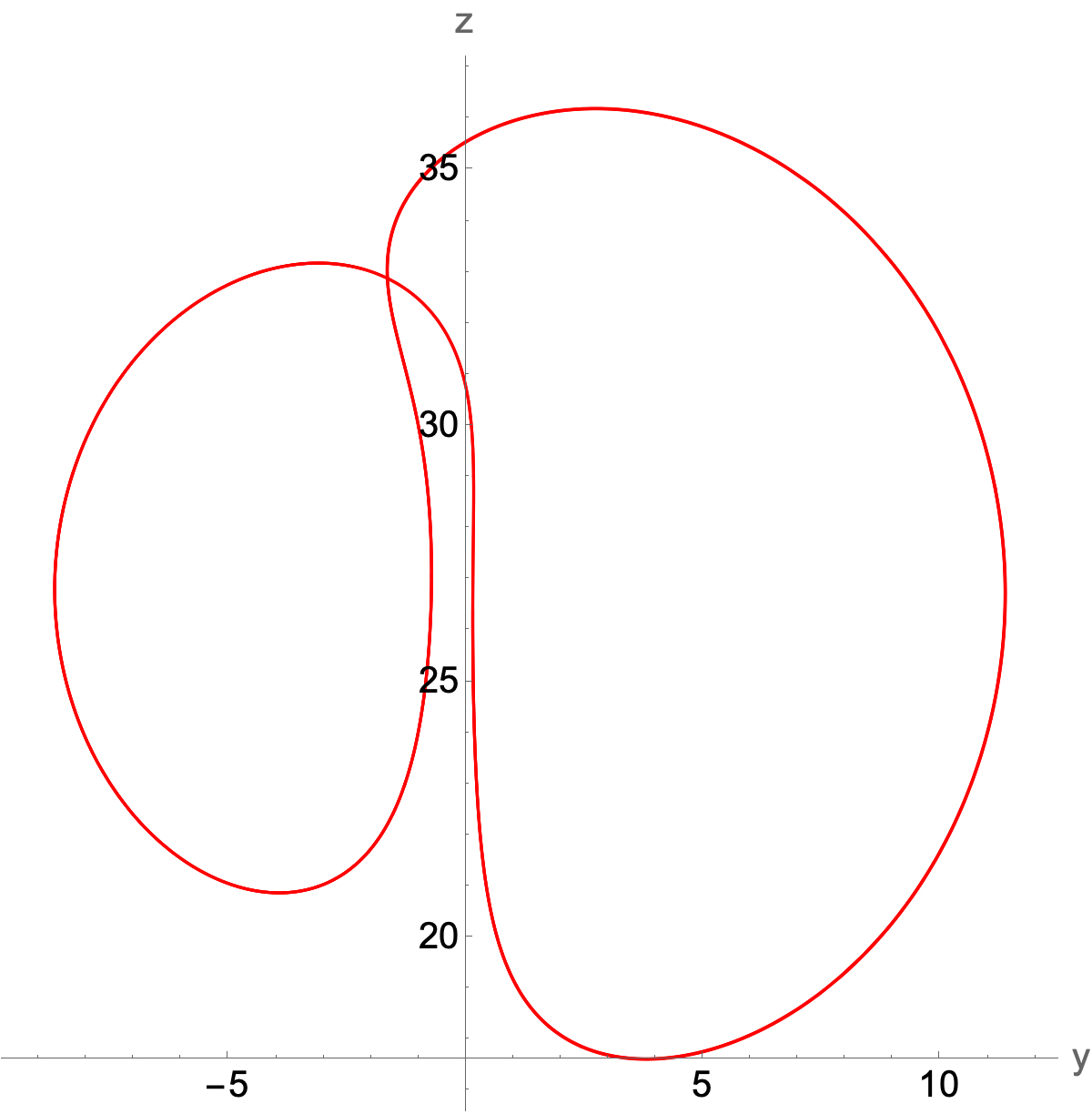}
  }
  \subfloat[\centering]{\includegraphics[width=0.32\linewidth]{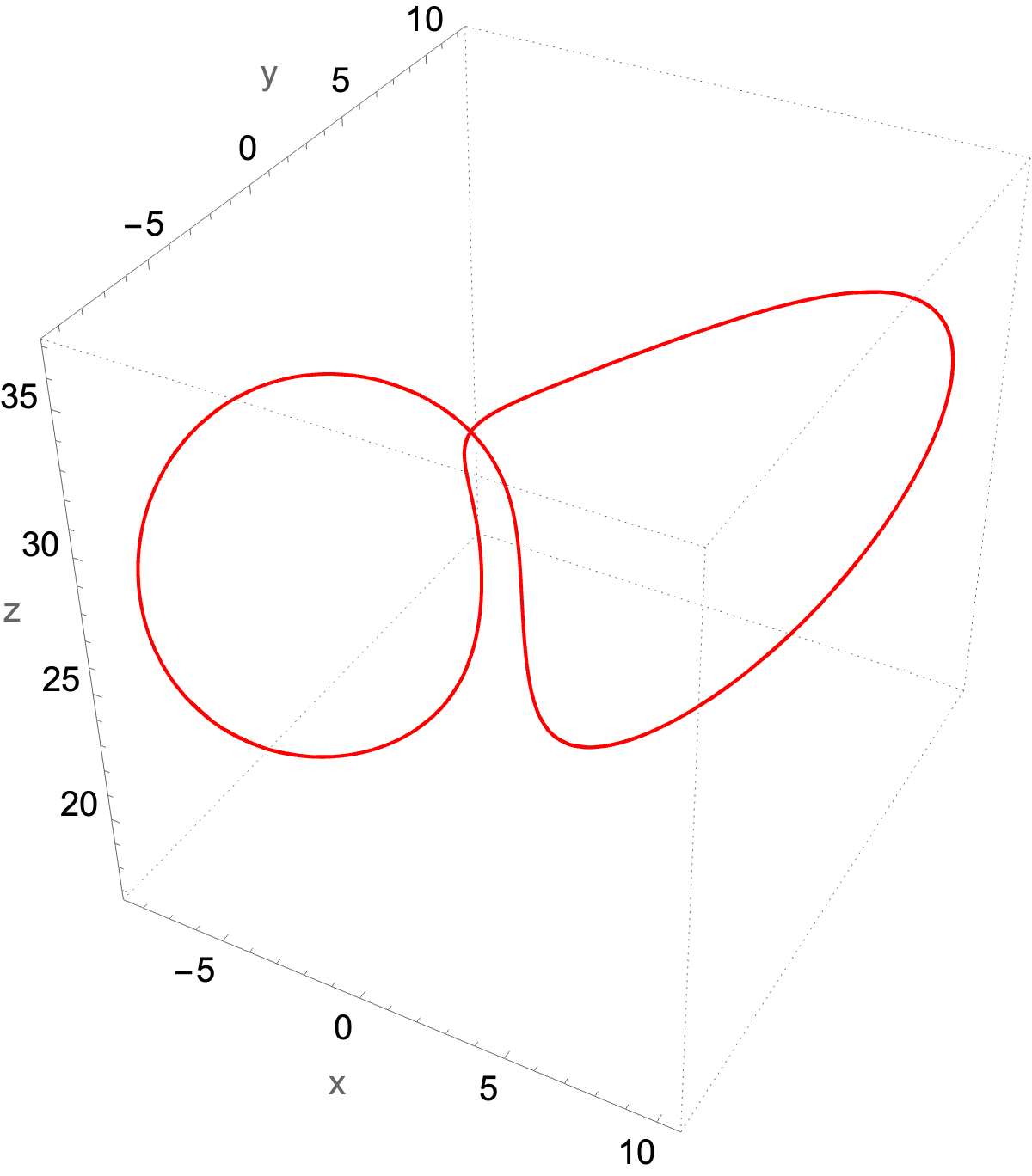}
  }
    \caption{$P1$-limit cycle with $\beta = 0.52$ and time step  $\delta t=10^{-3}$: (\textbf{a}) $y-z$ projection,  (\textbf{b})  the $3$D attractor.
}
    \label{p1 limit cycle}
\end{figure}

The period-$1$ limit cycle for $\beta = 0.52$ is shown in Figure~\ref{p1 limit cycle}, anchored about the two unstable fixed points $(\pm 3.75, \pm 3.75, 27)$.
\subsubsection{Period doubling limit cycles}
\begin{figure}[H]
\centering
\subfloat[\centering]{\includegraphics[width=0.35\linewidth]{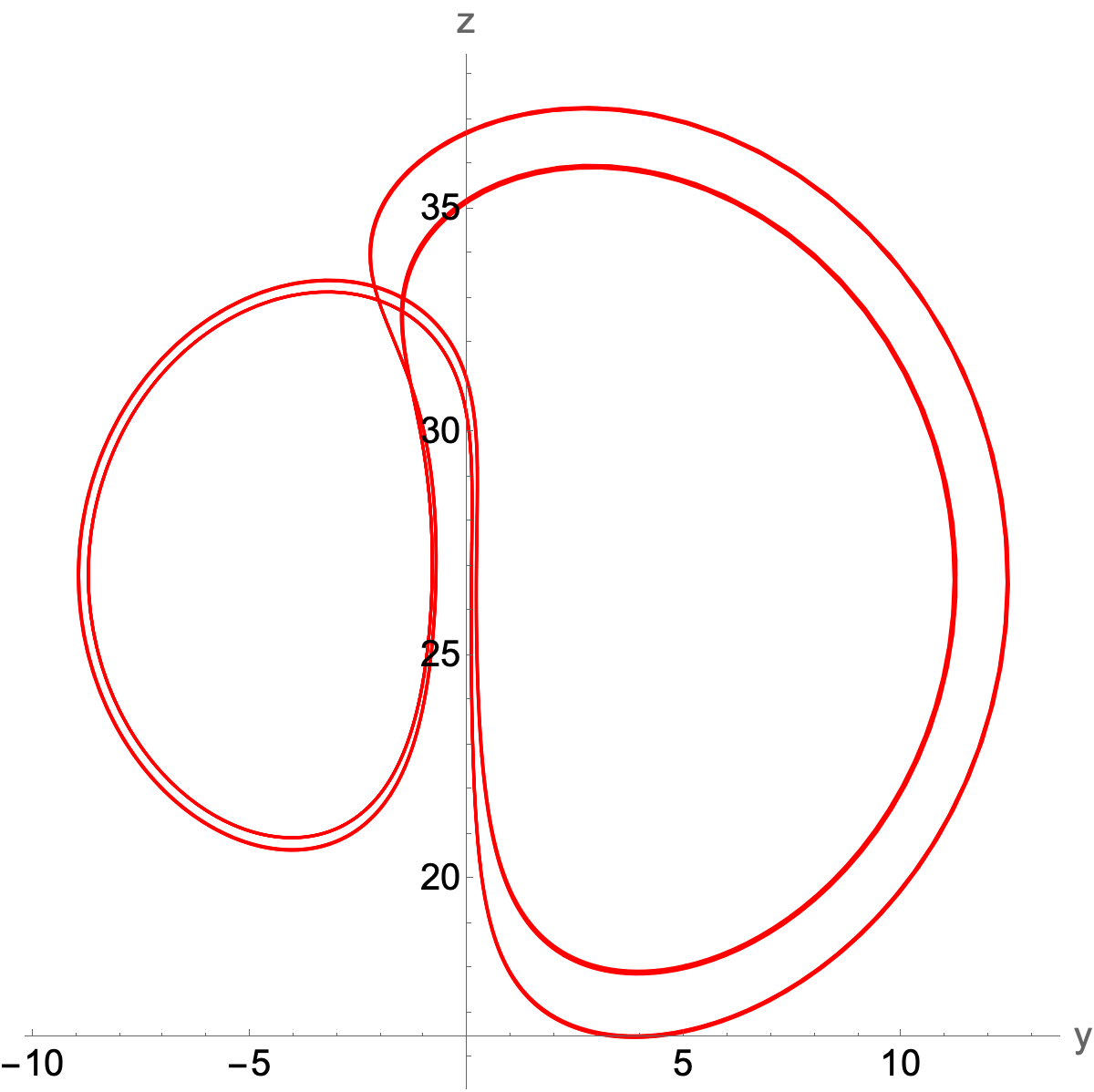}
  }
  \subfloat[\centering]{\includegraphics[width=0.32\linewidth]{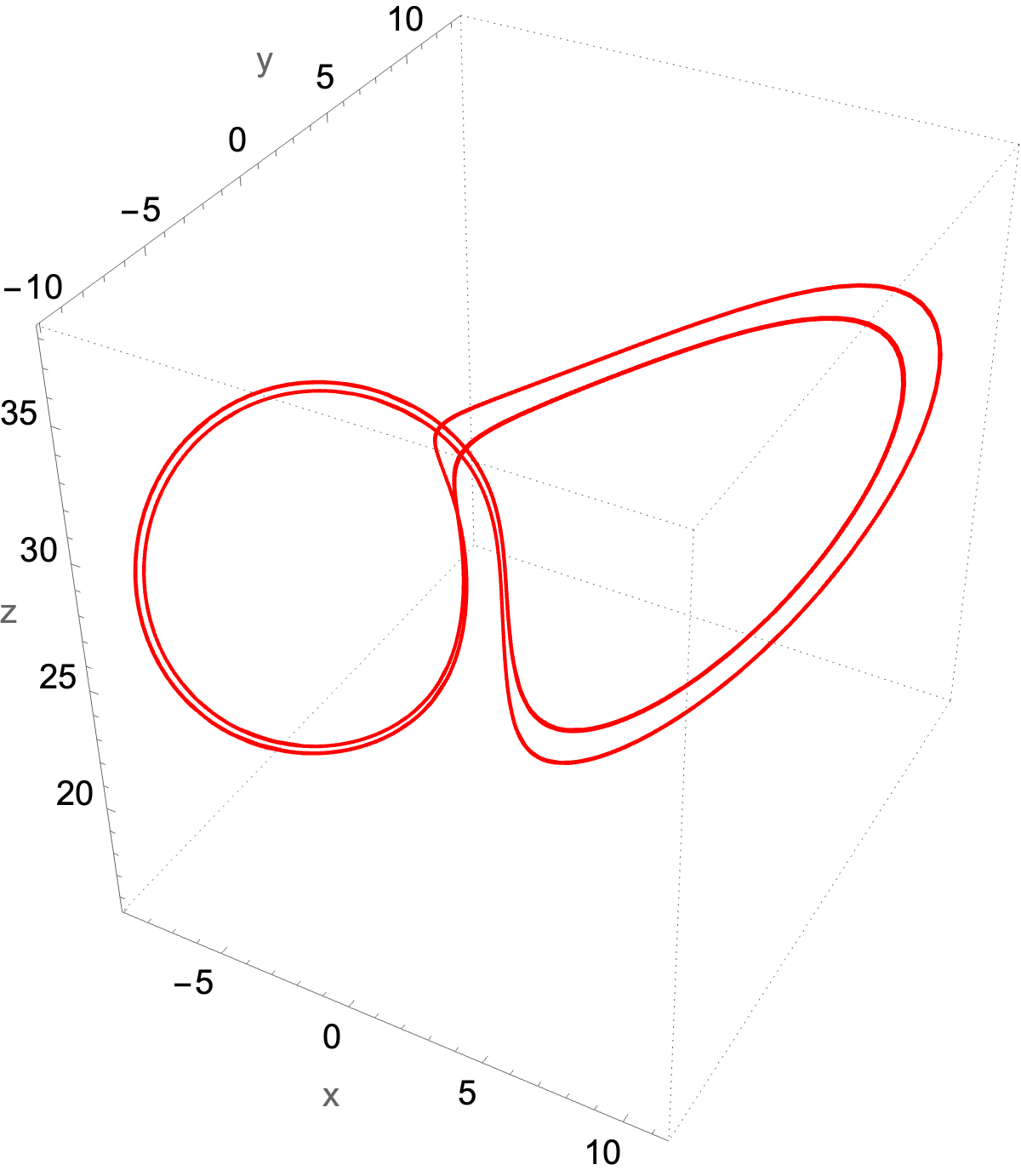}
  }
    \caption{$P2$-limit cycle, with $\beta = 0.55$ and time step  $\delta t=10^{-3}$: (\textbf{a}) $y-z$ projection,  (\textbf{b})  the $3$D attractor.
}
    \label{p2 limit cycle}
\end{figure}

\begin{figure}[H]
\centering
\subfloat[\centering]{\includegraphics[width=0.35\linewidth]{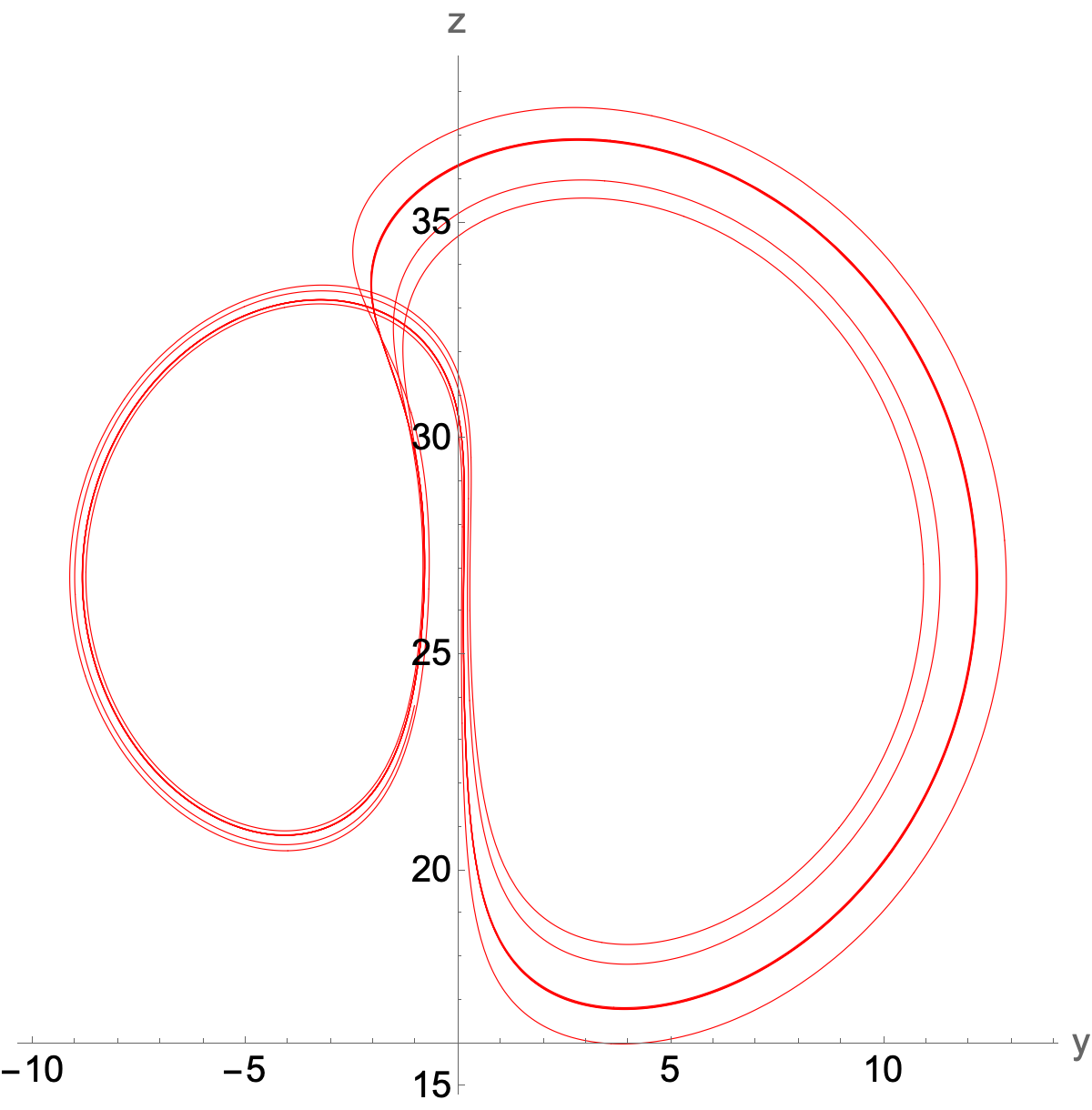}
  }
  \subfloat[\centering]{\includegraphics[width=0.32\linewidth]{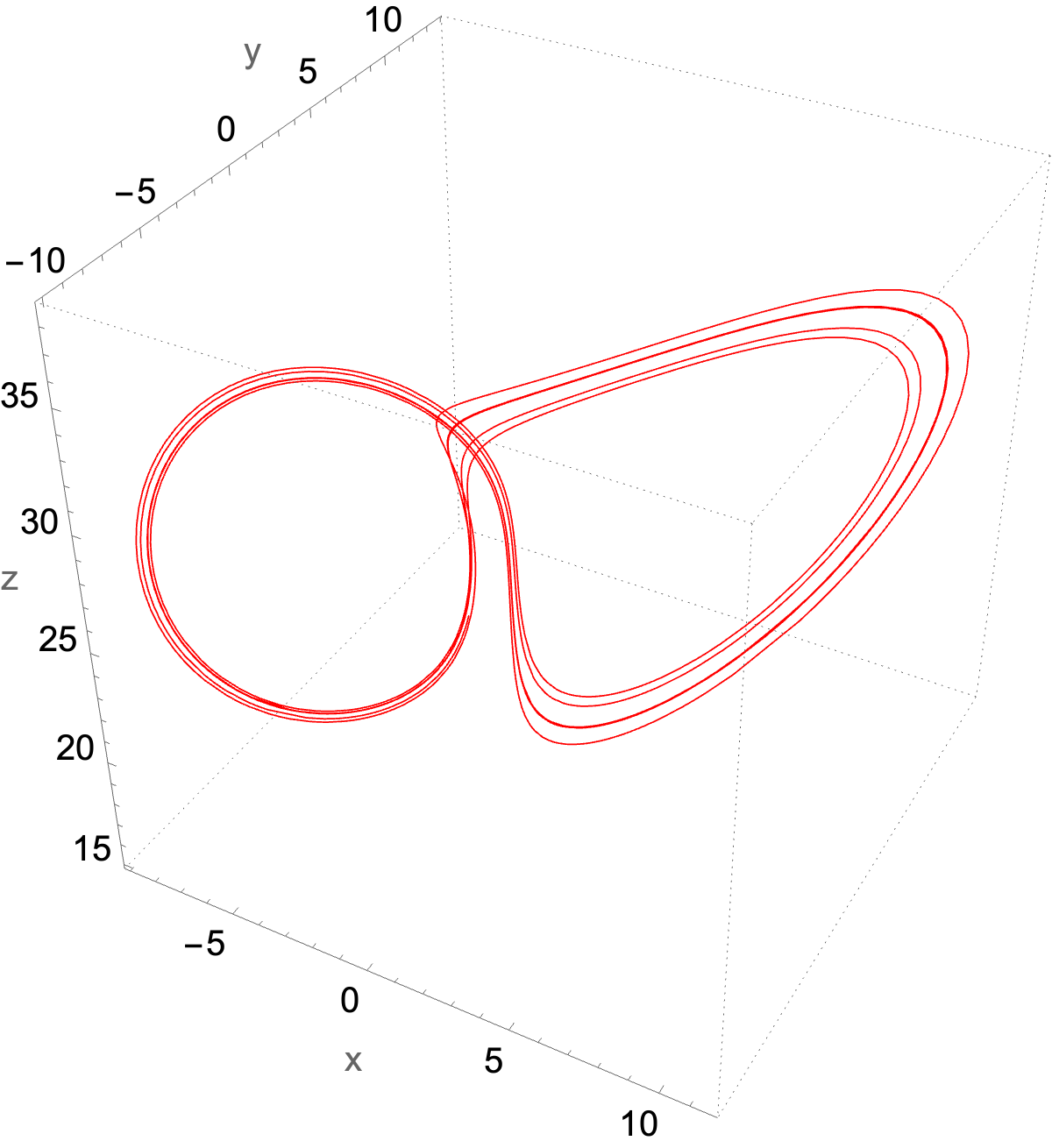}
  }
    \caption{$P2^2$-limit cycle, with $\beta = 0.56$ and time step  $\delta t=10^{-3}$: (\textbf{a}) y-z projection,  (\textbf{b}) the $3$D attractor.
}
    \label{p4 limit cycle}
\end{figure}

On increasing $\beta$, to $\beta = 0.55$ we encounter a period-2 $(P2)$ limit cycle, depicted in Figure~\ref{p2 limit cycle}, and a period-4 $(P2^2)$ limit cycle at $\beta = 0.56$  presented in Figure~\ref{p4 limit cycle} respectively.

The period-doubling limit cycles found in Figures~\ref{p1 limit cycle}-~\ref{p4 limit cycle} with the second-order method can be considered rudimentary in terms of their period-doubling behavior as the parameter $\beta$ increases towards the formation of the chaotic attractor. To further challenge the capabilities of the proposed second order quantum integration method we seek to verify whether we can generate a non trivial limit cycle. Such non-trivial limit cycles arise  as transitions from chaos to regulars dynamics as indicated in the bifurcation diagram.

\begin{figure}[H]
\centering
\subfloat[\centering]{\includegraphics[width=0.35\linewidth]{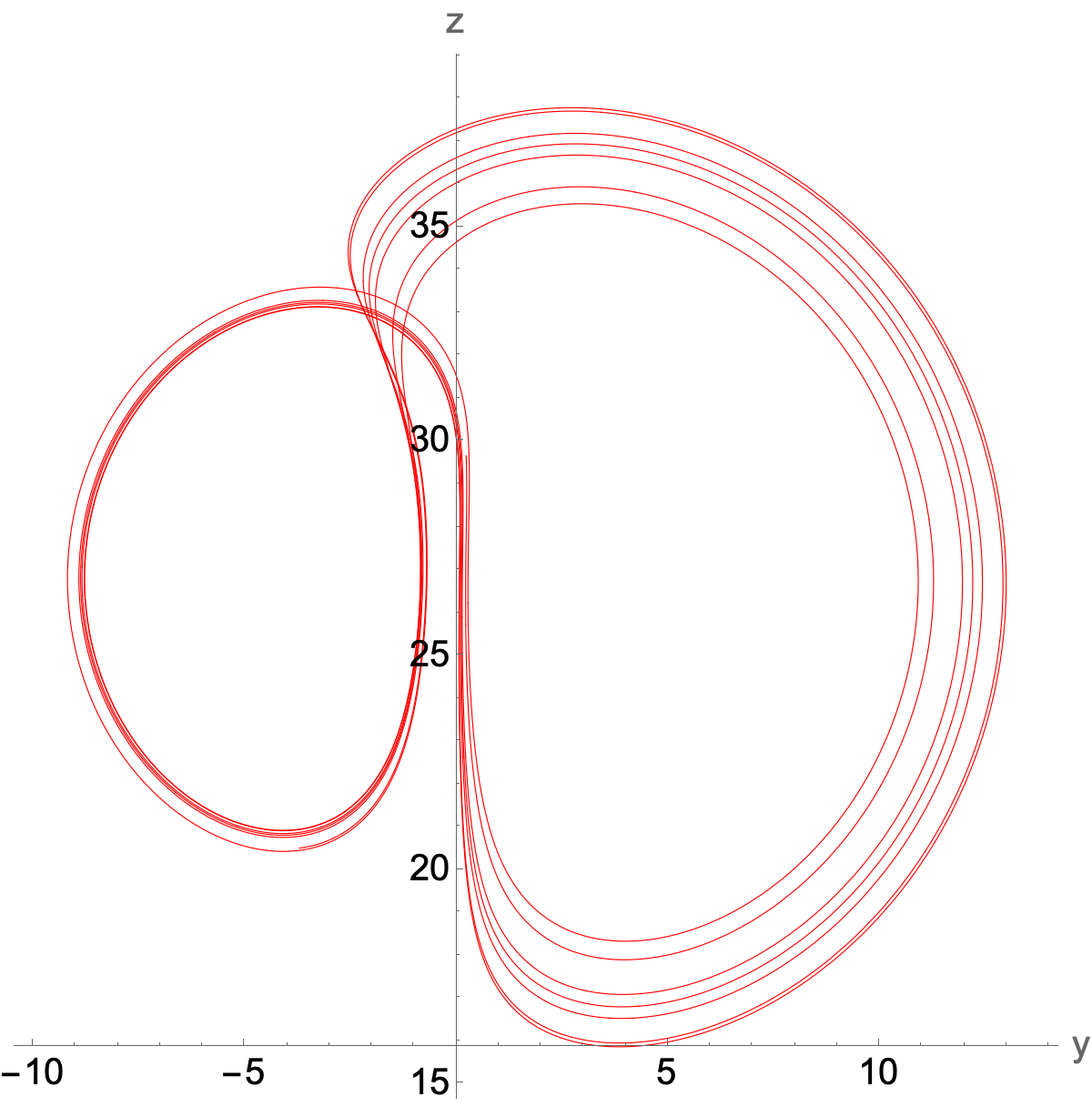}
  }
  \subfloat[\centering]{\includegraphics[width=0.32\linewidth]{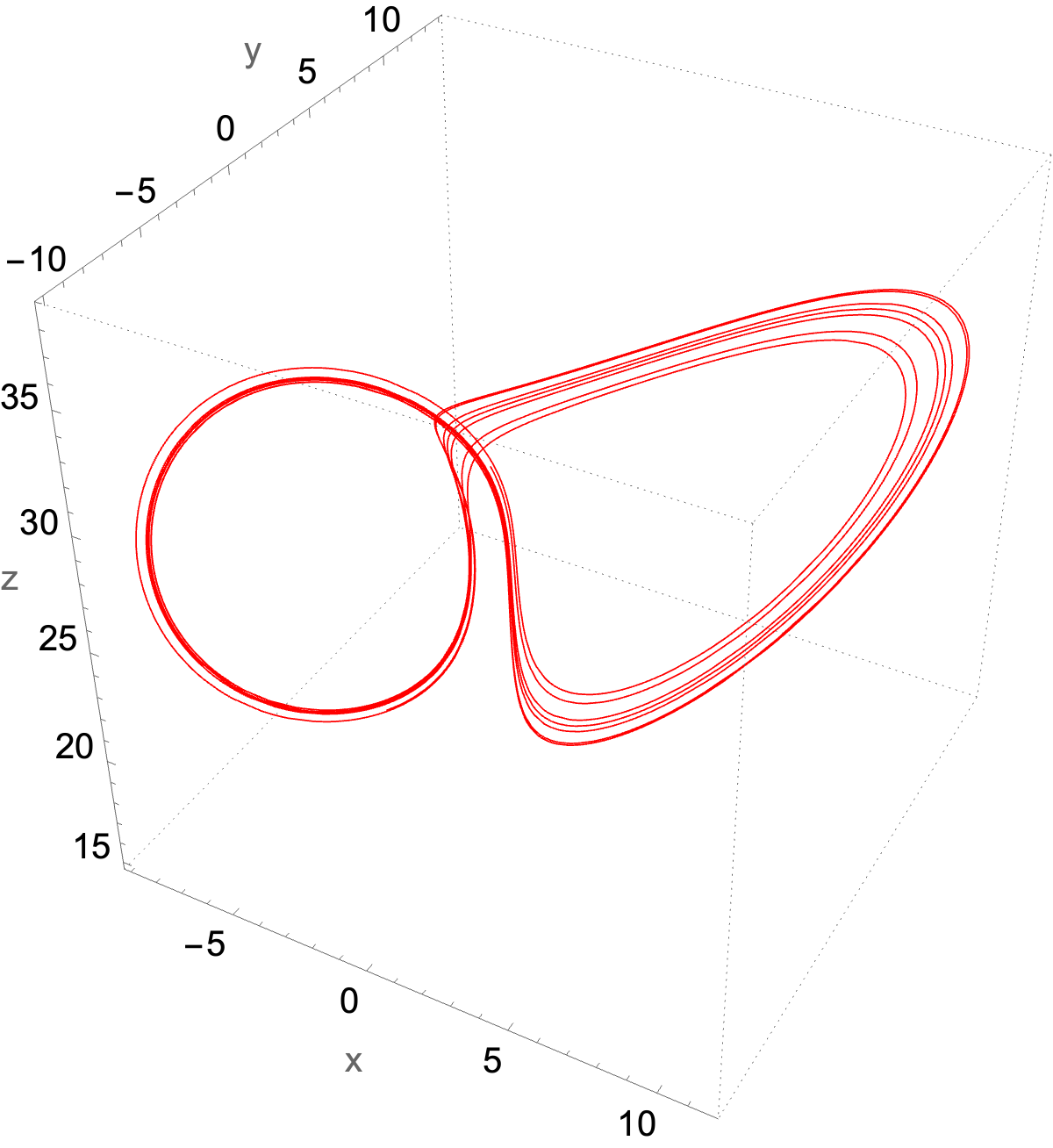}
  }
    \caption{$P6$-limit cycle for $\beta = 0.5648$ and a reduced time step $\delta t = 5 \cross 10^{-4}$ : (\textbf{a}) y-z projection,  (\textbf{b}) the $3$D attractor.
}
    \label{2nd order p6}
\end{figure}

\begin{figure}[H]
\centering
\subfloat[\centering]{\includegraphics[width=0.35\linewidth]{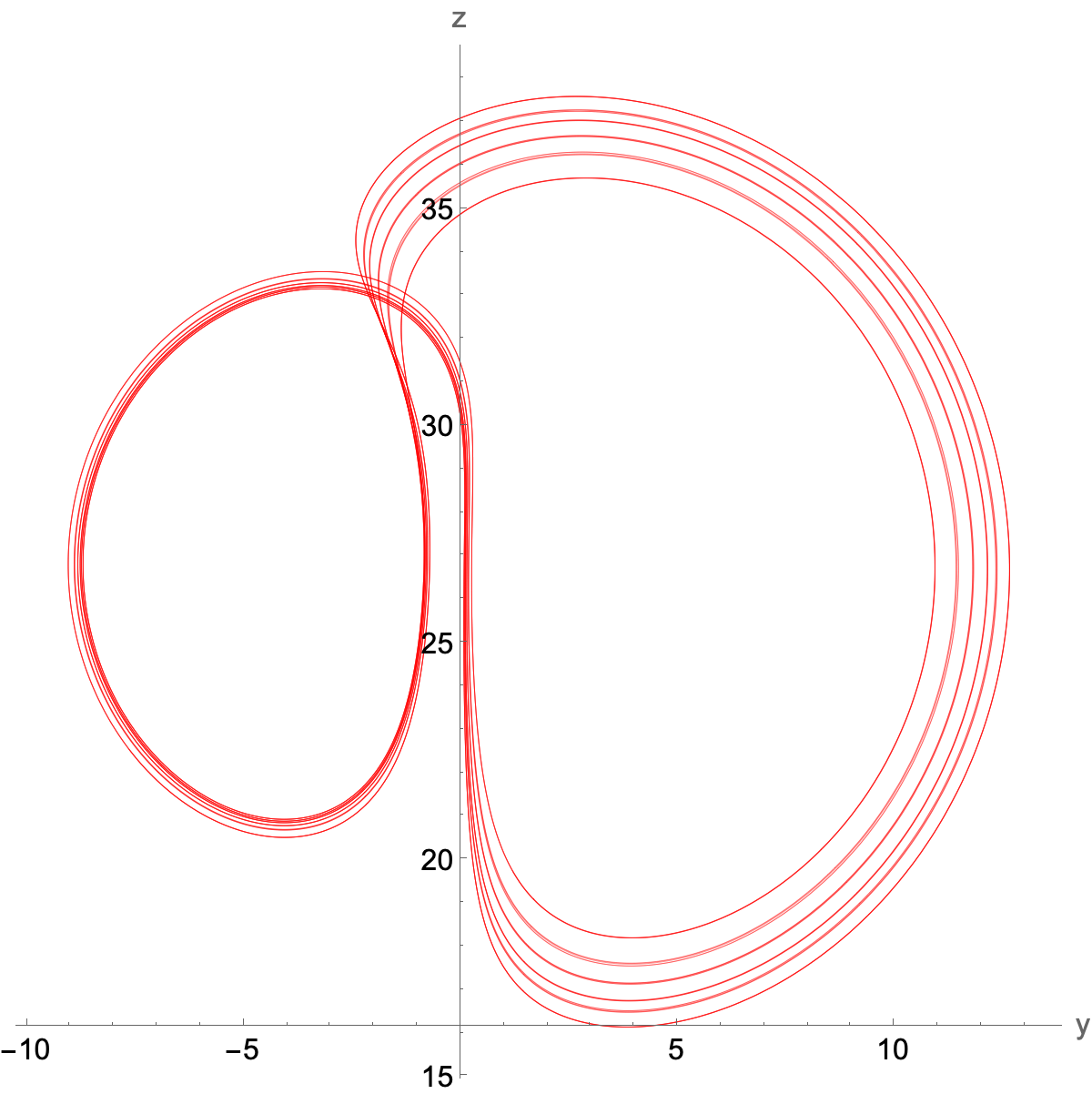}
  }
  \subfloat[\centering]{\includegraphics[width=0.32\linewidth]{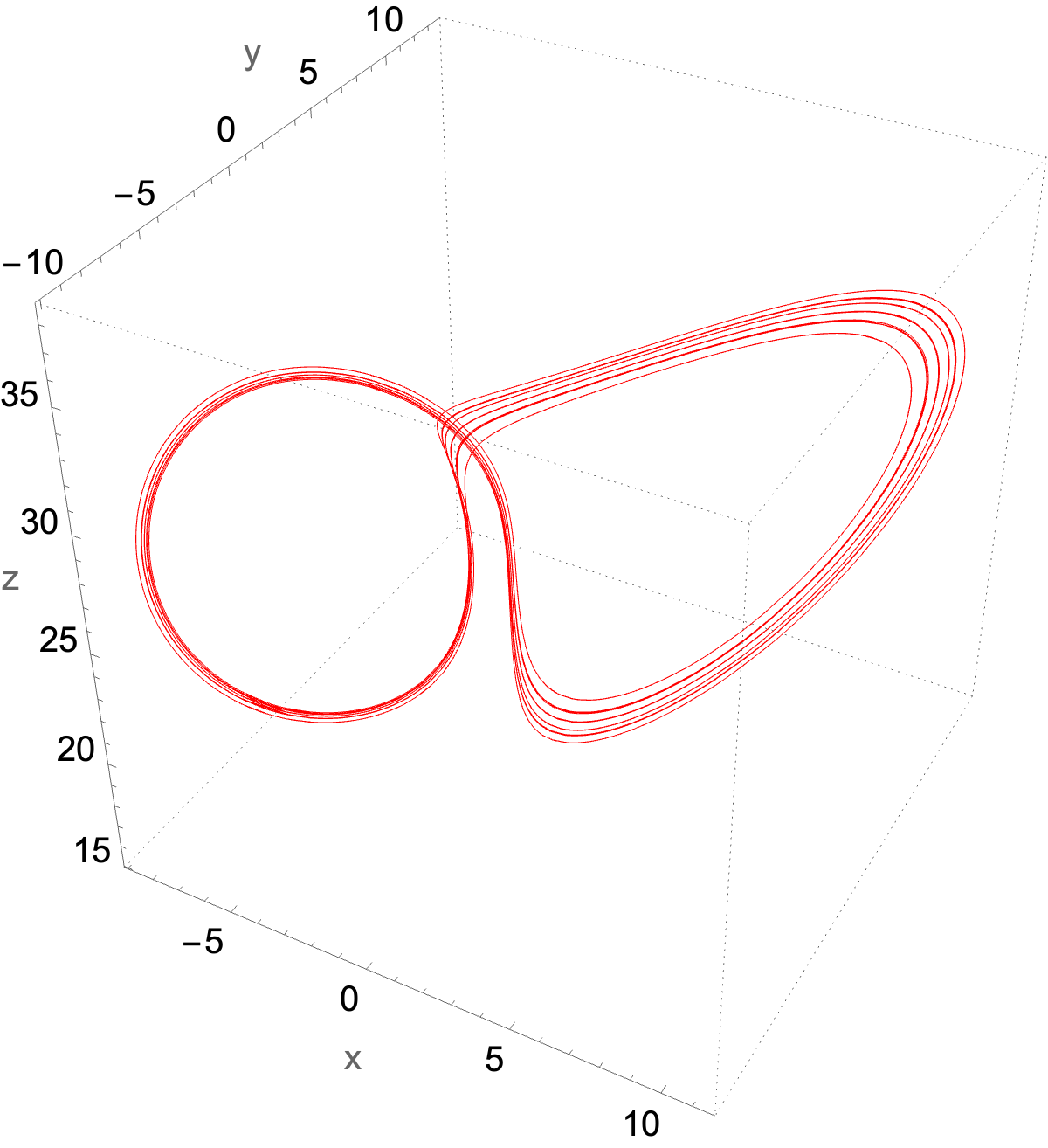}
  }
    \caption{$P6$-limit cycle generated by a higher order adaptive time $Mathematica$ ODE solver. (\textbf{a}) y-z projection,  (\textbf{b}) the $3$D attractor. Compared to the second-order scheme result in Figure~\ref{2nd order p6}, the main characteristics of the $P6$ structure are accurately captured.
}
    \label{Mathematica p6}
\end{figure}

For instance, at $\beta = 0.5648$, the Lorenz attractor is a $P6$ limit cycle resulting from the period-doubling bifurcation of the two $P3$ attractors. Such isolated and fine-structured solutions, cannot be obtained by series truncation techniques such as Carleman and KvN or by variational algorithms. In contrast, our method, accurately captures the structure of the $P6$-limit
cycle in Figure~\ref{2nd order p6} when compared to the $Mathematica$ ODE solver in Figure~\ref{Mathematica p6}.

\section{Conclusions}\label{sec:5}

In the present paper, we present a quantum algorithm that belongs to the class of quantum time-marching solvers, applied to a second-order discretized Lorenz system of nonlinear differential equations. For a single time step update, a two-stage process is required, first the preparation of a nonlinear quantum state and then the linear evolution of the state. To construct the nonlinear $4$-qubit Lorenz state $\ket*{\bol\psi_n^{nl}}$ the Hadamard product is employed whereas the action of the linear non-unitary operator $\hat{\mathcal{A}_2}$ is performed through a simple SVD-LCU block encoding technique. Extending the single-time step evolution into a complete time-marching scheme requires evolving multiple copies of the initial state $\ket{\boldsymbol{\psi}_n}$ in parallel, due to the no-cloning theorem. By incorporating  a re-usage of states~\citep{Esmaeilifar_2024} strategy, our algorithm obtains a recursive structure that allows for linear scaling in the number of state copies with respect to the number of iteration steps $N_t$ (Equation~\eqref{number of copies}). This represents an improvement over other quantum time-marching algorithms, which require exponential or quadratic resource scaling.

The algorithm can then be extended  to various classes of nonlinear differential equation systems, offering a quantum speed-up in respect to system's dimensionality $d_s$, as indicated in Equation~\eqref{total complexity}. Incidentally, the recursive structure  of the proposed algorithmic scheme has been identified as a promising feature for developing novel quantum algorithms with a quantum advantage~\citep{Bao_2024}.

Classical implementation of the proposed quantum algorithm for the Lorenz system demonstrates strong agreement with the $Mathematica$ generated chaotic and regular attractors. Combined with the improved quantum resources scaling presented in Equation~\eqref{total number of qubits}, these results suggest that the proposed time-marching quantum algorithm for discretized nonlinear differential equations has the potential to enable significant advances in the quantum simulation of classical nonlinear dynamics.

Additionally, it is of great interest to extend the presented ideas to partial differential equations (PDEs) employing a second order time discetization scheme but now within the framework of Qubit Lattice Algorithms (QLA). A Qubit lattice algorithm consists of a sequence of interleaved unitary collide-stream operators recovering the first order time discretized PDE  to a second order accuracy in spatial grid size $\delta x$. In the continuum limit, it recovers the correct partial differential equation provided the time displacement $\delta t$ obeys a diffusion ordering:  $\delta t \approx \delta x^2$.  This thus leads to an essentially second order QLA scheme. 

Qubit lattice algorithms have been applied to the nonlinear Gross–Pitaevskii system of equations, which describe the ground-state dynamics of spinor Bose–Einstein condensates~\citep{Vahala_2003,Vahala_2006,Yepez_2009,Zhang_2011,Vahala_2020}. In particular, in one dimension, QLA was used to study the long-time evolution of both scalar and vector soliton collisions. The numerical results were compared to exact analytical solutions of the nonlinear equations, demonstrating excellent agreement even though the time advancement
scheme appears to be a simple Euler scheme.

Finally,  QLA has been also employed for solving linear partial differential equations, such as Maxwell's equations, exhibiting a quantum speed-up compared to classical finite time difference schemes~\citep{Koukoutsis_2025b}.

\vspace{6pt} 





\authorcontributions{``Conceptualization, E.K. and G.V.;  methodology, E.K.; software, G.V., M.S. and L.V.; validation, K.H. and A.K.R.; formal analysis, E.K. and K.H.; investigation, E.K and G.V.; data curation, G.V., M.S. and L.V.; writing---original draft preparation, E.K. and G.V.; writing---review and editing, all authors have contributed equally; visualization, M.S. and L.V.; supervision, K.H. and A.K.R.;  funding acquisition,  all authors have contributed equally. All authors have read and agreed to the published version of the manuscript.''}

\funding{This work has been carried out within the framework
of the EUROfusion Consortium, funded by the European Union via the Euratom Research and Training Programme (Grant Agreement No 101052200 — EUROfusion). Views and opinions expressed are however those of the authors only and do not necessarily reflect those of
the European Union or the European Commission. Neither the European Union nor the European Commission can be held responsible for them. A.K.R., G.V., M.S. and
L.V. are supported by the US Department of Energy under Grant Nos. DE-SC0021647, DE-FG02-91ER-54109,
DE-SC0021651, DE-SC0021857 and DE-SC0021653.}

\institutionalreview{Not applicable.}



\dataavailability{The data generated in this study are available upon reasonable request from the authors.}

\acknowledgments{E.K acknowledges valuable discussions on quantum computing techniques and quantum algorithms with {\'O}scar Amaro and Lucas I. I{\~n}igo Gamiz and thanks them for reviewing and commenting on the original draft. Additionally, E.K and K.H acknowledge discussions with Ioannis Theodonis, Christos Tsironis, Panagiotis Papagiannis, Yannis Kominis, Aristides Papadopoulos, Elias Glytsis and Giorgos Fikioris.}

\conflictsofinterest{The authors declare no conflicts of interest. The funders had no role in the design of the study; in the collection, analyses, or interpretation of data; in the writing of the manuscript; or in the decision to publish the results.} 



\abbreviations{Abbreviations}{
The following abbreviations are used in this manuscript:
\\

\noindent 
\begin{tabular}{@{}ll}
KvN & Koopman von Neumann\\
CNOT & Controlled-NOT\\
LCU & Linear Combination of Unitaries\\
SVD & Singular Value Decomposition\\
ODE & Ordinary differential equation\\
NISQ & Noisy Intermediate-Scale Quantum\\
PDE& Partial differential equation\\
QLA & Qubit Lattice Algorithm
\end{tabular}
}



\reftitle{References}


\bibliography{ref}


%


\PublishersNote{}
\end{document}